\documentclass[aps,prd,reprint,showpacs,showkeys,floatfix,amsmath,amssymb,nofootinbib,longbibliography,tabulary,color]{revtex4-1}
\usepackage{amsmath,amssymb,graphics,setspace}
\usepackage{bm,soul}
\usepackage[pdftex]{graphicx}
\usepackage{multirow}
\usepackage{color}
\usepackage{courier}
\usepackage{xspace}
\usepackage{tabularx,ragged2e,booktabs}
\usepackage{array}

\usepackage[usenames,dvipsnames,svgnames,table]{xcolor}
\usepackage{natbib}
\usepackage{hyperref} 
\hypersetup{colorlinks=true,linkcolor=blue,filecolor=magenta,urlcolor=red,citecolor=blue}

\usepackage{dcolumn}
\usepackage{tabulary}
\usepackage{siunitx}
\usepackage{longtable}
\usepackage{enumitem}
\usepackage{url}

\newcommand{\ten}[2]{\ensuremath{{#1}{\times} 10^{#2}}}

\newcommand{\olam}{\Omega_{\Lambda}}
\newcommand{\om}{\Omega_{\rm m}}
\newcommand{\orad}{\Omega_{\rm r}}
\newcommand{\ok}{\Omega_{\rm k}}

\newcommand{\Gev}{{\rm \, GeV}}

\newcommand{\tnot}{t_0}
\newcommand{\Ho}{H_0}

\newcommand{\Hounits}{km s$^{-1}$Mpc$^{-1}$}

\newcommand{\Ip}{I_{p}}

\newcommand{\Pq}{q}
\newcommand{\Pu}{u}

\newcommand{\stwo}{S5 B0716+714}

\newcommand{\Yjm}{\ensuremath{Y_{jm}}}
\newcommand{\Yjmi}{\ensuremath{Y_{jm,{\sourceindex}}}}
\newcommand{\Yjma}{\ensuremath{\Yjm(\theta_{\sourceindex},\phi_{\sourceindex})}}
\newcommand{\oYjm}{   \ensuremath{ {}_{0}\Yjm }    }

\newcommand{\twoYjmi}{   \ensuremath{ {}_{\pm2}\Yjmi }    }
\newcommand{\cdIjm}{\ensuremath{\smash{c^{(d)}_{(I)jm}}}}
\newcommand{\kdVjm}{\ensuremath{\smash{k^{(d)}_{(V)jm}}}}
\newcommand{\cdI}{\ensuremath{\smash{\bar{c}^{(d)}_{(I),{\sourceindex}}}}}
\newcommand{\cdIo}{\ensuremath{\smash{\bar{c}^{(d)}_{(I)}}}}

\newcommand{\cdIsixo}{\ensuremath{\smash{\bar{c}^{(6)}_{(I)}}}}
\newcommand{\cdIeighto}{\ensuremath{\smash{\bar{c}^{(8)}_{(I)}}}}

\newcommand{\kdV}{\ensuremath{\smash{\bar{k}^{(d)}_{(V),{\sourceindex}}}}}
\newcommand{\kdVo}{\ensuremath{\smash{\bar{k}^{(d)}_{(V)}}}}
\newcommand{\cdIoo}{\ensuremath{\smash{c^{(d)}_{(I)00}}}}
\newcommand{\csixIoo}{\ensuremath{\smash{c^{(6)}_{(I)00}}}}

\newcommand{\kdVoo}{\ensuremath{\smash{k^{(d)}_{(V)00}}}}
\newcommand{\kdVfive}{\ensuremath{\smash{\bar{k}^{(5)}_{(V),{\sourceindex}}}}}

\newcommand{\kdVfiveo}{\ensuremath{\smash{\bar{k}^{(5)}_{(V)}}}}
\newcommand{\kdVseveno}{\ensuremath{\smash{\bar{k}^{(7)}_{(V)}}}}
\newcommand{\kdVnineo}{\ensuremath{\smash{\bar{k}^{(9)}_{(V)}}}}
\newcommand{\kdEB}{\ensuremath{\smash{\bar{k}^{(d)}_{(EB),{\sourceindex}}}}}

\newcommand{\kdEBo}{\ensuremath{\smash{\bar{k}^{(d)}_{(EB)}}}}

\newcommand{\kdEBsix}{\ensuremath{\smash{\bar{k}^{(6)}_{(EB),{\sourceindex}}}}}

\newcommand{\kdEBfouro}{\ensuremath{\smash{\bar{k}^{(4)}_{(EB)}}}}
\newcommand{\kdEBsixo}{\ensuremath{\smash{\bar{k}^{(6)}_{(EB)}}}}
\newcommand{\kdEBeighto}{\ensuremath{\smash{\bar{k}^{(8)}_{(EB)}}}}

\newcommand{\kdEjm}{\ensuremath{\smash{k^{(d)}_{(E)jm}}}}
\newcommand{\kdBjm}{\ensuremath{\smash{k^{(d)}_{(B)jm}}}}

\newcommand{\sourceindex}{{s}}

\newcommand{\zi}{(z_{\sourceindex})}

\newcommand{\Dtd}{\Delta t^{(d)}_{\zi}}

\newcommand{\zz}{z}
\newcommand{\aaa}{a}

\newcommand{\Ldz}{\smash{L^{(d)}_{\zi}}}
\newcommand{\Pqdz}{\smash{q^{(d)}_{\zi}}}
\newcommand{\Pudz}{\smash{u^{(d)}_{\zi}}}
\newcommand{\Pmaxdz}{\smash{p_{max,\zi}^{(d)}}}
\newcommand{\pmax}{p_{max}}

\newcommand{\Ldzfive}{\smash{L^{(5)}_{\zi}}}
\newcommand{\Ldzsix}{\smash{L^{(6)}_{\zi}}}
\newcommand{\zetakfive}{ \smash{\zeta_{\sourceindex}^{(5)}}}
\newcommand{\zetaksix}{ \smash{\zeta_{\sourceindex}^{(6)}}}

\newcommand{\cosphiavg}{ \langle \cos \Phi_{\sourceindex} \rangle }

\newcommand{\cspace}{\hspace{0.9em}}

\newcommand{\pstaroneLumbllac}{$1.01 \pm 0.03$}
\newcommand{\pstartwoLumbllac}{$0.92 \pm 0.07$}
\newcommand{\pstaroneIcbllac}{$0.55 \pm 0.03$}
\newcommand{\pstartwoIcbllac}{$0.46 \pm 0.07$}

\newcommand{\pstaroneLumsfive}{$0.26 \pm 0.43$}
\newcommand{\pstartwoLumsfive}{$0.21 \pm 0.27$} 
\newcommand{\pstaroneIcsfive}{$0.89 \pm 0.73$}
\newcommand{\pstartwoIcsfive}{$0.77 \pm 0.23$}

\newcommand{\tdelaybllac}{$26.5 \pm 19.5$}
\newcommand{\tdelaysfive}{$-5.1 \pm 3.3$}
\newcommand{\tdelayupbllac}{65.5}
\newcommand{\tdelayupsfive}{11.7}

\newcommand{\pstar}{p_{\star}}
\newcommand{\pisp}{p_{\rm sys,ISP}}
\newcommand{\pinst}{p_{\rm sys,inst}}
\newcommand{\pstarcor}{p_{\star,cor}}

\newcommand{\apol}{APPOL}

\newcommand{\orcidwidth}{0.1in} 

\renewcommand{\prl}{Phys. Rev. Lett.}
\renewcommand{\rmp}{Rev. Mod. Phys.}
\newcommand{\plb}{Phys. Lett. B}

\newcommand{\pasp}{Publ. Astron. Soc. Pac.}
\newcommand{\mnras}{Mon. Not. R. Astron Soc.}

\newcommand{\aap}{Astron. Astrophys.}
\newcommand{\jcap}{J. Cos. Astropart. Phys.}
\newcommand{\apjl}{Astrophys. J. Lett.}
\newcommand{\apjs}{Astrophys. J. Suppl. Ser.}
\newcommand{\pasj}{Publ. Astron. Soc. Jap.}
\newcommand{\aj}{Astron. J.}
\newcommand{\pasa}{Publ. Astron. Soc. Aus.}
\newcommand{\physrep}{Phys. Rep.}
\newcommand{\app}{Astropart. Phys.}

\newcommand{\aspcs}{Astron. Soc. Pac. Conf. Ser.}

\newcommand{\ascl}{Astrophys. Source Code Lib.}

\newcommand{\sasas}{Soc. Astron. Sci. Ann. Symp.}

\newcommand{\aipcs}{Amer. Inst. Phys. Conf. Ser.}

\newcommand{\jpcs}{Jour. Phys. Conf. Ser.}

\newcommand{\grg}{Gen. Rel. Grav.}

\newcommand{\mpla}{Mod. Phys. Lett. A}

\newcommand{\ijmpd}{Int. Jour. Mod. Phys. D}

\newcommand{\cpc}{Chin. Phys. C}

\newcommand{\natp}{Nature Phys.}

\newcommand{\procspie}{Proc. SPIE}

\newcommand{\aapr}{Astron. Astrophys. Rev.}

\newcommand{\jqsrt}{Jour. Quant. Spec. Rad. Trans.}

\newcommand{\actaa}{Acta Astronomica}

\newcommand{\aasma}{Am. Astron. Soc. Meet. Abs.}

\newcommand{\jaavso}{Jour. Am. Assoc. Var. Star Obs.}

\newcommand{\al}{Astron. Lett.}

\newcommand{\aag}{Astron. Geophys.}

\begin{document}

\title{Constraints on Lorentz Invariance and CPT Violation using Optical Photometry and Polarimetry of Active Galaxies BL Lacertae and \stwo}

\author{Andrew~S.~Friedman 
\href{https://orcid.org/0000-0003-1334-039X}{\includegraphics[width=\orcidwidth]{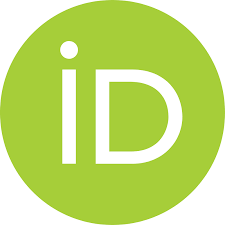}}}
\email{asf@ucsd.edu}
\affiliation{Center for Astrophysics and Space Sciences, University of California, San Diego, La Jolla, California 92093, USA}

\author{Gary~M.~Cole
\href{https://orcid.org/0000-0002-8041-5649}{\includegraphics[width=\orcidwidth]{./figs/orcid.png}}}
\email{garycole@mac.com}
\affiliation{Starphysics Observatory, 14280 W. Windriver Lane, Reno, NV 89511}

\author{David~Leon}
\email{dleon@physics.ucsd.edu}
\affiliation{Center for Astrophysics and Space Sciences, University of California, San Diego, La Jolla, California 92093, USA}

\author{Kevin~D.~Crowley}
\affiliation{Center for Astrophysics and Space Sciences, University of California, San Diego, La Jolla, California 92093, USA}

\author{Delwin~Johnson}
\affiliation{Center for Astrophysics and Space Sciences, University of California, San Diego, La Jolla, California 92093, USA}

\author{Grant~Teply}
\affiliation{Center for Astrophysics and Space Sciences, University of California, San Diego, La Jolla, California 92093, USA}

\author{David~Tytler}
\affiliation{Center for Astrophysics and Space Sciences, University of California, San Diego, La Jolla, California 92093, USA}

\author{Brian~G.~Keating 
\href{https://orcid.org/0000-0003-3118-5514}{\includegraphics[width=\orcidwidth]{./figs/orcid.png}}}
\email{bkeating@ucsd.edu}
\affiliation{Center for Astrophysics and Space Sciences, University of California, San Diego, La Jolla, California 92093, USA}

\date{\today}

\begin{abstract}
Various quantum gravity approaches that extend beyond the standard model predict Lorentz Invariance and Charge-Parity-Time Violation at energies approaching the Planck scale. These models frequently predict a wavelength dependent speed of light, which would result in time delays between promptly emitted photons at different energies, as well as a wavelength-dependent rotation of the plane of linear polarization for photons resulting from vacuum birefringence. Here, we describe a pilot program with an automated system of small telescopes that can simultaneously conduct high cadence optical photometry and polarimetry of Active Galactic Nuclei (AGN) in multiple passbands. We use these observations as a proof-of-principle to demonstrate how such data can be used to test various Lorentz Violation models, including special cases of the Standard Model Extension (SME). In our initial campaign with this system, the Array Photo Polarimeter, we observed two AGN sources, including BL Lacertae at redshift $z=0.069$, and \stwo{} at $z=0.31$. We demonstrate that optical polarimetry with a broadband $Luminance$ filter combined with simultaneous $I_c$-band observations yields SME parameter constraints that are up to $\sim$10 and $\sim$30 times more sensitive than with a standard $I_c$-band filter, for SME models with mass dimension $d=5$ and $d=6$, respectively. Using only a small system of telescopes with an effective $0.45$-m aperture, we further demonstrate $d=5$ constraints for individual lines of sight that are within a factor of $\sim$$1$-$10$ in sensitivity to comparable constraints from optical polarimetry with a $3.6$-m telescope. Such an approach could significantly improve existing
SME constraints via a polarimetric all-sky survey of AGN with multiple 1-meter class telescopes.
\end{abstract}


\maketitle

\section{Introduction}
\label{sec:intro}

Special relativity and the standard model of particle physics obey the symmetry of Lorentz Invariance, which has survived an enormous range of tests over the past century (See \cite{kostelecky11} for a review). However, many theoretical approaches seeking to unify quantum theory and general relativity predict that Lorentz Invariance may be broken at energies approaching the Planck scale $E_p = \sqrt{c^5 \hbar / G} = \ten{1.22}{19} \Gev$, perhaps due to the underlying quantized nature of spacetime (e.g. \cite{myers03,amelino15}). Since the relevant energies are not accessible to any current, or foreseeable, Earth-bound tests, most approaches to testing such models have relied on observations of high redshift astronomical sources to exploit small effects that may accumulate to detectable levels over cosmological distances and timescales. 

This paper considers only Lorentz Invariance Violation (LIV) for photons\footnote{Other authors have considered testing LIV models for massive particles including neutrinos, which can be considered as approximately massless  \cite{jacob07,jacob08,chakraborty13,stecker14,amelino15} and cosmic rays \cite{scully09,stecker10,bietenholz11,cowsik12,lang17}.}, which can lead to a modified vacuum dispersion relation, and therefore an energy dependent speed of light, which causes a time delay (or early arrival) for promptly emitted photons of different energies \cite{jacob08,kostelecky08}. LIV models can also yield vacuum birefringence, which causes a rotation of the plane of linear polarization for promptly emitted photons at different energies emitted with the same initial polarization angle \cite{carroll90,kostelecky08}. In general, each of these effects can be anisotropic, such that time delays and polarization rotations possess an angular dependence on the sky, and require observations of extended sources like the Cosmic Microwave Background (CMB) or measurements of point sources along many lines of sight to fully test the LIV model parameter space \cite{kostelecky09,kislat17}. 

\begin{table*}[t]
\centering 
\begin{tabular}{cccccccccc} 
\hline \hline
  \multicolumn{1}{c}{Name}               
& \multicolumn{1}{c}{RA}    
& \multicolumn{1}{c}{DEC} 
& \multicolumn{1}{c}{Redshift $z$}  
& \multicolumn{1}{c}{$z$ Ref.}  
& \multicolumn{1}{c}{$B$}    
& \multicolumn{1}{c}{$V$} 
& \multicolumn{1}{c}{$R$}
& \multicolumn{1}{c}{$Lum$} 
& \multicolumn{1}{c}{$I_c$} \\ 
  \multicolumn{1}{c}{}        
& \multicolumn{1}{c}{IRCS(J2000)$^{\circ}$}   
& \multicolumn{1}{c}{IRCS(J2000)$^{\circ}$}    
& \multicolumn{1}{c}{} 
& \multicolumn{1}{c}{} 
& \multicolumn{1}{c}{(mag)} 
& \multicolumn{1}{c}{(mag)} 
& \multicolumn{1}{c}{(mag)} 
& \multicolumn{1}{c}{(mag)} 
& \multicolumn{1}{c}{(mag)} \\ 
\hline 
S5 0716+714 &	110.47270192  &	+71.34343428   &    0.31 $\pm$	0.08 & \cite{nilsson08,danforth13}  &	15.50 &	14.17 &	14.27 & 14.65 & 14.10	\\
BL Lacertae & 	330.68038079 &	+42.27777231   &    0.0686	$\pm$ 0.0004 & \cite{vermeulen95} &	15.66  &	14.72	& 13.00 & 13.89 & 13.06 \\
\hline 
\hline 
\end{tabular} 
\caption{
Celestial coordinates and $BVR$ magnitudes of observed AGN sources from the Simbad database (Magnitudes may not be typical for these variable sources). $Lum$ and $I_c$ magnitudes are mean values from our own photometry in Tables~\ref{tab:data1}-\ref{tab:data2}. 
}
\label{tab:params} 
\end{table*} 
\renewcommand{\arraystretch}{1} 

Testing LIV is difficult because each of these effects are expected to be negligible at energies accessible in Earth-bound or solar system experiments. However, any such effects could, in principle, accrue to measurable levels as these tiny deviations from Lorentz symmetry accumulate over cosmological distances. Qualitatively, evidence for LIV time delays from photometric observations are easier to measure for sources at higher cosmological redshifts and higher energies \cite{jacob07,jacob08,kostelecky08,kostelecky09}. Compared to time delays, birefringent LIV models can be tested with much higher sensitivity using spectropolarimetry or broadband polarimetry \cite{kostelecky09}. 

In this work, we restrict our analysis to constraining a subset of the Standard Model Extension (SME), an effective field theory approach describing the low energy corrections stemming from a more fundamental Planck scale theory of quantum gravity. The SME therefore provides a general framework for Lorentz Invariance and Charge-Parity-Time (CPT) violation tests with electromagnetic radiation \cite{kostelecky09}.\footnote{We therefore do not consider models such as Doubly (or Deformed) Special Relativity (e.g. \cite{amelino10,smolin11}), which may not be compatible with the SME \cite{kostelecky09,kislat17}.}
More specifically, since we are only reporting observations of two optical AGN sources, we are limited to constraining either general SME models along specific lines of sight or vacuum isotropic SME models, which correspond to some of the more popular 
models studied in the literature. We further confine our analysis to SME models of mass operator dimension $d=4,5,6,7,8,9$. Mass dimension $d=3$ models are best constrained with observations of the CMB \cite{kostelecky08,kostelecky09,komatsu09,gubitosi09,kahniashvili09,kaufman16a,leon17}\footnote{For a discussion of the difficulties in calibrating the reference angle for astrophysical CMB polarization measurements, see \cite{kaufman16b} and \cite{kaufman14}.}. While $d=4$ models can yield birefringent effects, they would not produce LIV induced time delays, since they involve no changes to the usual photon dispersion relation \cite{kostelecky09}.

Simultaneous photometric observations in two filters allows one to estimate upper limits to time delays between light curves in each bandpass.  While our optical time delay constraints are not competitive with observations of gamma-ray bursts \cite{amelino98,boggs04,ellis06,rodriguez06,kahniashvili06,biesiada07,xiao09,laurent11,stecker10,stecker11,toma12,kostelecky13,vasileiou13,pan15,zhangs15,chang16,lin16,wei17} or TeV flares from blazars \cite{biller99,albert08,aharonian08,kostelecky08,shao10,tavecchio16}, our approach, which may be unique in the literature, does constrain both time delays and maximum observed polarization with simultaneously obtained photometry and polarimetry using the same pair of broadband optical filters. As such, they have the promise to compliment existing SME constraints.

Time delay measurements uniquely constrain the SME vacuum dispersion coefficients and, in principle, could constrain the vacuum birefringent coefficents as well for all models with $d \ne 4$.  However, time delay measurements are typically less sensitive than broadband polarimetry for constraining the birefringent SME coefficients \cite{kostelecky09}, so we exclusively use broadband polarimetry to constrain all other SME coefficients. While optical spectropolarimetry can yield constraints $\sim$$2$-$3$ orders of magnitude better for $d=5$ models than broadband optical polarimetry \cite{kislat17}, this generally requires $\gtrsim$ 2-meter class telescopes. With telescopes less than 1-m in diameter, broadband polarimetry in two or more filters is considerably more practical, offering a solution that is low-cost and scalable to large numbers of observatories around the world. Since we did not obtain spectropolarimetry in our pilot program, we focus on the broadband polarimetry method for the rest of this work.

When observing a single source, as noted by \cite{ellis06}, it is, in general, impossible to disentangle an intrinsic time-lag at the source from a delay induced by genuine LIV dispersion effects.\footnote{Note that the cosmological time delay calculation from \cite{ellis06} contains a basic error which was noted and corrected by \cite{jacob08} and used by subsequent analyses (e.g. \cite{kostelecky08,kislat17}). This issue is also relevant for LIV tests using gravitational lensing \cite{biesiada09} and pulsar timing \cite{shao14}.} Therefore, to constrain LIV models using observed time delays, one must either assume A) there are no intrinsic time delays, or B) statistically model observations of many sources using the fact that all LIV effects are predicted to increase with redshift and therefore be negligible for sufficiently ``nearby" sources. For approach B, one would model the population distribution of intrinsic time lags using a calibration sample of low redshift sources and use this to disentangle these non-LIV effects from genuine LIV effects which could be manifest in a suitably matched population of higher redshift sources \cite{ellis06,kostelecky09,kislat17}.  However, since we only observed one nearby source (BL Lacertae at $z = 0.0686 \pm 0.0004$; \cite{vermeulen95}) and one high redshift source (\stwo{} at $z=0.31 \pm 0.08$, \cite{nilsson08,danforth13}), we assume option A for the remainder of this work. 

Similarly, it is, in general, impossible to know the intrinsic polarization angles for photons emitted with different energies from a given cosmological source. If one possessed this information, evidence for birefringence could be obtained by observing differences between the known intrinsic polarization angle and the actual observed angles for photons emitted promptly with the same polarization angle but at different energies. However, even in the absence of such knowledge, birefringent effects can be constrained for sources at arbitrary redshifts because a large degree of birefringence would yield large differences in observed polarization angles at nearby frequencies, effectively washing out most, if not all, of the observed polarization \cite{kostelecky09,kostelecky13,lin16}. Therefore, observing a given polarization fraction can constrain wavelength-dependent birefringence effects, which, if in effect, would have led to a smaller degree of observed polarization. To analyze SME models in this work, we follow the ``average polarization" approach in \cite{kislat17}.\footnote{The authors in \cite{kislat17} also analyzed both optical polarimetry and spectropolarimetry, where available, from 72 existing polarized AGN and Gamma-Ray Burst (GRB) afterglow sources in the literature (e.g. \cite{schmidt92,sluse05,smith09}).}

In this work, we present simultaneous photometric and polarimetric observations using two broadband optical filters on separate telescopes, including the $Luminance$-band filter ($Lum$) and a Johnson-Cousins $I$-band filter ($I_c$). 
While not as common as standard optical $BVRI$ filters, we chose the wider $Lum$ filter both to maximize the signal for our small telescopes and because wider optical bandpasses lead to tighter constraints on birefringent SME Models obtained using any of the standard optical $BVRI$ filters \cite{kislat17}. In particular, we demonstrate significant advantages of the wider $Lum$ filter versus the narrower $I_c$ filter, where, for the same observed maximum polarization fraction, the $Lum$ filter yields $d=5,6$ SME parameter upper bounds that are factors of $\sim$$3$-$26$ times more sensitive than with the $I_c$-band filter. 

In addition, we develop a technique to combine simultaneous polarimetric observations using two co-located telescopes with different filters into an effective system with a single broadband optical filter that avoids the expense of a half-wave plate with high transmission over the full $\sim$400-900\,nm wavelength range of the combined $Lum+I_c$ filter. This yields more stringent SME constraints than either filter alone, while achieving the effective light collecting power of a larger telescope. This approach can be contrasted with an optical system using dichroic beamsplitters on a single, large telescope, to obtain simultaneous polarimetry in different bandpasses (e.g. the DIPOL-2 instrument \cite{piirola14}). With this approach, for the same observed maximum polarization fraction, our combined $Lum+I_c$ filter yields $d=5,6$ SME parameter upper bounds that are factors of $\sim$$2$-$30$ times more sensitive than with the $I_c$-band filter.

The pilot program in this work is meant as a proof-of-principle to obtain the most stringent SME constraints using broadband optical polarimetric observations with small telescopes for which spectropolarimetry is unfeasible. Even without spectropolarimetry, anisotropic SME constraints can be improved by observing sources along lines of sight without previously published optical polarimetry. Even if specific AGN sources already have published optical polarimetry, improved SME constraints can potentially be obtained simply by observing these sources with wider optical bandpasses, and by potentially observing a larger maximum polarization value than previously found. For all of these reasons, this work aims to motivate design feasibility studies for a follow up optical polarimetry survey using at least two 1-m class telescopes, with one or more in each hemisphere. 

This paper is organized as follows. In \S\ref{sec:sme}, we describe the Standard Model Extension family of Lorentz and CPT-Invariance violating models we are interested in testing and present our main constraints. In \S\ref{sec:apol}, we describe the optical polarimetric and photometric observing systems used in this work, with emphasis on correcting for systematic errors in our maximum polarization measurements. Conclusions are presented in \ref{sec:conc}. Mathematical details, and the data obtained for this paper are presented in the Appendix.

\section{Standard Model Extension}
\label{sec:sme}

We do not describe the full SME framework here. Instead, see \cite{kostelecky09} for a review. Qualitatively, if the Standard Model holds perfectly, all SME coefficients vanish identically. No strong evidence yet exists for any non-zero SME coefficients, and therefore, many LIV models falling under the SME umbrella have already been ruled out. However, the general approach to make progress testing such models is to use observations of cosmological sources at different wavelengths, higher redshifts, and varied positions on the sky to progressively lower the upper bounds for any non-zero values of the coefficients over the full SME parameter space. Weak constraints imply very large, uninformative, upper bounds. Strong constraints imply very small, informative, upper bounds that constrain coefficient values progressively closer to zero. However, even seemingly weak constraints can be of value of they are obtained with an observational approach with smaller (or different) systematics than an approach that nominally yields stronger constraints \cite{kislat17}. 

\subsection{Vacuum Dispersion SME Models}
\label{sec:smevacdis}

Most LIV models predict a wavelength-dependent speed of light, leading to light of a given energy arriving earlier (or later) than light of another energy, even if both were emitted simultaneously in the rest frame of the source. Following \cite{jacob08,kostelecky08,kostelecky09,kostelecky13}, in the context of the SME, the arrival time difference between photons emitted simultaneously from a cosmological source with index label $\sourceindex$ at redshift $z=z_{\sourceindex}$ and sky position $(\theta_{\sourceindex},\phi_{\sourceindex})$, with observed energies $E_1$ and $E_2$, (and detected at observer frame times $t_1$ and $t_2$, respectively), is given by
\begin{eqnarray}
\Dtd & =  & t_2 {\scriptstyle -} t_1  \approx \left(E_2^{d-4}{\scriptstyle -}E_1^{d-4}\right) \frac{\Ldz}{c}
\sum_{jm} \Yjmi \cdIjm\, , \nonumber \\
\label{eq:deltat}
\end{eqnarray}
where $\Yjmi \equiv \Yjma$ are the spin weighted spherical harmonics for spin-0\footnote{$\Yjm \equiv \oYjm$ are the usual spherical harmonics for spin-0.}, $\cdIjm$ are the vacuum dispersion SME coefficients with mass dimension $d=4,6,8,\ldots$ which must be CPT-even, and
\begin{eqnarray}
\frac{\Ldz}{c} = \int_{0}^{z_{\sourceindex}} \frac{ (1+\zz)^{d-4} }{ H(\zz) }d\zz =  \int_{a_{\sourceindex}}^{1} \frac{d\aaa}{ (\aaa)^{d-2} H(\aaa)} \, ,
\label{eq:Lz}
\end{eqnarray}
where $\Ldz$ is the effective comoving distance traveled by the photons, including the cosmological effects needed to compute arrival time differences in an expanding universe \cite{jacob08}. Setting $d=4$ recovers the usual expression for comoving distance. In Eq.~(\ref{eq:Lz}), $H(z)=H(a)$ is the Hubble expansion rate at a redshift $z_{\sourceindex}$ with scale factor $a_{\sourceindex}^{-1}=1+z_{\sourceindex}$ (with the usual normalization $a(\tnot)=1$ at the present cosmic time $t=\tnot$ at $z=0$) given by
\begin{eqnarray}
H(a) = \Ho \Big[\orad a^{-4} + \om a^{-3} + \ok a^{-2} + \olam \Big]^{1/2}\, ,
\label{eq:Hz}
\end{eqnarray}
in terms of the present day Hubble constant, which we set to $\Ho=73.24$ \Hounits{} \cite{riess16}, and best fit cosmological parameters for matter $\om = 0.3089$, radiation $\orad = \om/(1+z_{eq}) =\ten{9.16}{-5}$ (with the matter-radiation equality redshift $z_{eq}=3371$), vacuum energy $\olam = 0.6911$, and curvature $\ok = 1 - \orad - \om - \olam \approx 0$ using the Planck satellite 2015 data release \cite{planck16}.\footnote{We use cosmological parameters reported in Table 4 column 6 of \cite{planck16}. These are the joint cosmological constraints (TT,TE,EE+lowP+lensing+ext 68\% limits (where ext=BAO+JLA+H0)). However, based on recent tension between the Hubble constant $\Ho$ determined using CMB data and Type Ia supernovae (SN Ia), we use the SN Ia Hubble constant $\Ho=73.24$ \Hounits \cite{riess16} rather than $\Ho=67.74$ \Hounits from Table 4 coLumn 6 of \cite{planck16}.}

In principle, observations constraining the theoretical time delay $\Dtd$ from Eq.~(\ref{eq:deltat}) between photons observed at different energies can constrain the SME coefficients $\cdIjm$. More specifically, an upper bound $|\Delta t_{\star}|$ on the theoretical time delay (or early arrival)  $|\Dtd| \le |\Delta t_{\star}|$ measured from photometry in different bandpasses can be recast as an upper bound on a linear combination of SME coefficients:
\begin{eqnarray}
\cdI \equiv \Big| \sum_{jm} \Yjmi \cdIjm \Big| & \lesssim & \frac{ c\, |\Delta t_{\star}|}{\left|E_2^{d-4}{\scriptstyle -}E_1^{d-4}\right| \Ldz}\, ,
\label{eq:deltatup}
\end{eqnarray}
where $E_1$ and $E_2$ can be estimated from the central wavelengths of the filters. Eq.~(\ref{eq:deltatup}) defines $\cdI$ as shorthand for the absolute value of the linear combination of vacuum dispersion SME coefficients for source $\sourceindex$.

\begin{figure}
\begin{tabular}{@{}c@{}}

\includegraphics[width=3.5in]{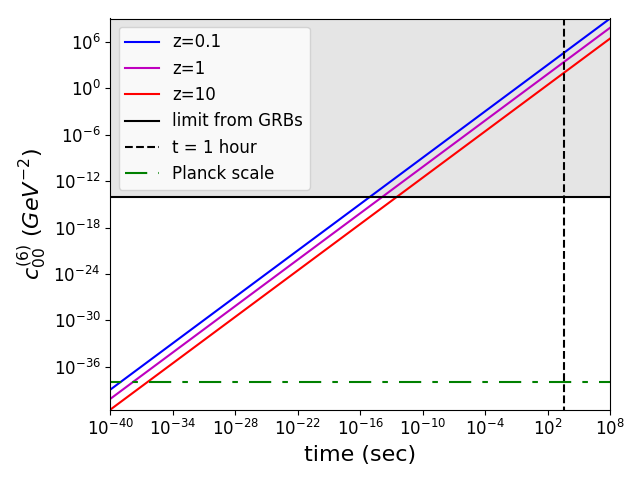} \\

\end{tabular}
\caption{
\baselineskip 9pt 
We plot the dimension $d=6$ isotropic vacuum dispersion SME parameter $\csixIoo$ for time delays between two example observations in the $Lum$ and $I_c$ bands (central wavelengths of $\sim$550\, nm vs. $\sim$800\, nm) for various redshift sources ($z=0.1,1,10$). The horizontal dot-dashed line shows the $\csixIoo$ corresponding to the Planck energy scale, while the dashed vertical line corresponds to a time delay of 1 hour. Gray regions in the parameter space with $\csixIoo \lesssim 10^{-14}$ GeV$^{-2}$ have already been ruled out by high redshift, high time resolution, Gamma-Ray burst data \cite{kostelecky13}. Because of this, optical time delays on the order of minutes to hours for moderate redshift sources can only provide weak --- but still independent --- constraints as a consistency check.
}
\label{fig:opticalbandSME}
\end{figure}

Fig.~\ref{fig:opticalbandSME} shows the relation between time delay upper limits and $d=6$ isotropic SME models for sample sources observed with both our $Lum$ and $I_c$ filters over a range of redshifts $z \in [0.1,1,10]$, while highlighting the parameter space already ruled out by limits from GRB observations, as well as the weaker, but meaningful constraints obtainable from optical time delay data with $|\Delta t_{\star}| \le 1$ hour.

\subsection{CPT-Odd Vacuum Birefringent SME Models}
\label{sec:smebiref}

For a subset of vacuum birefringent SME models with coefficients $\kdVjm$, where the mass dimension $d=3,5,7,\ldots$ must be CPT-odd, rather than arrival times, the relevant quantity is the rotation of the plane of linear polarization for photons with different observed energies $E_1$ and $E_2$ that were emitted in the rest frame of the source with the same polarization angle. After traveling an effective distance of $\Ldz$ through an expanding universe, the difference in their observed polarization angles $\Delta \psi^{(d)}_{\zi} = \psi_2 {\scriptstyle -} \psi_1$ will be
\begin{eqnarray}
\Delta \psi^{(d)}_{\zi} & \approx  &  \left(E_2^{d-3}{\scriptstyle -}E_1^{d-3}\right) \frac{\Ldz}{c} \sum_{jm} \Yjmi \kdVjm \, .
\label{eq:deltapsi}
\end{eqnarray}
In principle, polarimetric observations measuring an observed polarization angle difference $|\Delta \psi_{\star}|$
in a single broadband filter with bandpass edge energies $E_1$ and $E_2$, with $|\Delta \psi^{(d)}_{\zi}| \le |\Delta \psi_{\star}|$, can constrain the SME coefficients $\kdVjm$ directly using Eq.~(\ref{eq:deltapsi}), 
\begin{eqnarray}
\kdV \equiv \Big| \sum_{jm} \Yjmi \kdVjm \Big| & \le  & \frac{c\, |\Delta \psi_{\star}|}{\Big|E_2^{d-3}{\scriptstyle -}E_1^{d-3}\Big| \Ldz }\, ,
\label{eq:deltapsistar}
\end{eqnarray}
where $\kdV$ is shorthand for the absolute value of the linear combination of birefringent SME coefficients for source $\sourceindex$.\footnote{We present constraints from our data using the $Lum$ and $I_c$-band optical filters in Sec~\ref{sec:constraints}.}

Eq.~(\ref{eq:deltapsistar}) requires the assumption that all photons in the observed bandpass were emitted with the same (unknown) intrinsic polarization angle. When not making such an assumption, a more complicated and indirect argument is required.  In general, when integrating over an energy range $[E_1,E_2]$, if LIV effects exist, the observed polarization degree will be substantially suppressed for a given observed energy if $\Delta \psi_{\star} > \pi$, regardless of the intrinsic polarization fraction at the corresponding rest frame energy \cite{kostelecky13,kislat17}. Other authors present arguements allowing them to assume $\Delta \psi_{\star} \leq \pi/2$ to derive bounds on certain SME models \cite{toma12}. In our case, observing a polarization fraction $\pstar$ can be used to constrain birefringent SME coefficients as follows.

\begin{figure}
\centering
\begin{tabular}{@{}c@{}}

\includegraphics[width=3.5in]{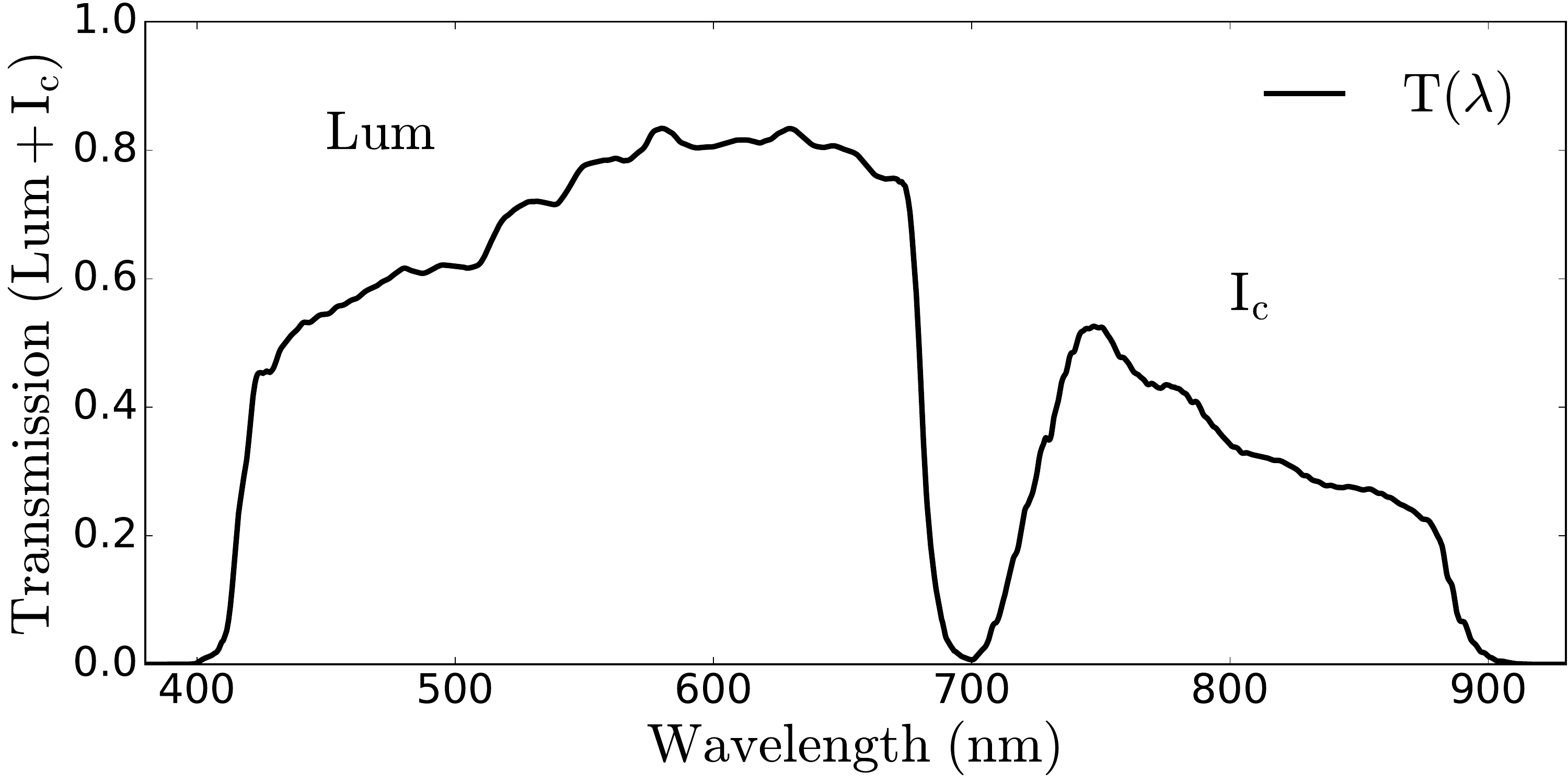} \\

\end{tabular}
\caption{
\baselineskip 9pt 
Total transmission function from optics, filters, and CCD detectors for our $Lum$ and $I_c$-bands observed using the Array Photo Polarimeter (\apol, see \S\ref{sec:apol}), which we combine into a single, effective broadband $Lum+I_c$ filter with coverage from $\sim$400-900 nm (with minimal filter overlap at $\sim$700 nm), using simultaneous data from two telescopes (see Fig.~\ref{fig:apoltrans}).
}
\label{fig:Lum+I_TE}
\end{figure}

First, one conservatively assumes a 100\% intrinsic polarization fraction at the source for all wavelengths. Lower fractions for the source polarization spectrum would lead to tighter SME bounds. In this case, the total intensity $I$ is equal to the polarized intensity $\Ip$, such that
\begin{eqnarray}
I = \int_{E_1}^{E_2} T(E) dE = \Ip\, ,
\label{eq:ITE}
\end{eqnarray}
where $T(E)$ is the total throughput transmission function as a function of photon energy $E=hc/\lambda$ (with wavelength $\lambda$) for the polarimeter, including the relevant optics, broadband filters, and detectors (see Fig.~\ref{fig:Lum+I_TE}). Then, following \cite{kislat17}, integrating Eq.~(\ref{eq:deltapsi}) over the energy range of the effective bandpass $T(E)$ yields normalized linear polarization Stokes parameters $\Pq \equiv Q/I$ and $\Pu \equiv U/I$, given by
\begin{eqnarray}
&& \Pqdz  =  \frac{\Ip}{I} \int_{E_1}^{E_2} \cos \Big( 2 \Delta \psi \Big) T(E) dE \\
 && = \int_{E_1}^{E_2} \cos \Big( 2 \left(E^{d-3}{\scriptstyle -}E_1^{d-3}\right) \Ldz \kdV \Big) T(E) dE \, ,\nonumber 
\label{eq:PqLIV}
\end{eqnarray}
and
\begin{eqnarray}
&& \Pudz  =  \frac{\Ip}{I} \int_{E_1}^{E_2} \sin \Big( 2 \Delta \psi \Big) T(E) dE \\
 && = \int_{E_1}^{E_2} \sin \Big( 2 \left(E^{d-3}{\scriptstyle -}E_1^{d-3}\right) \Ldz \kdV \Big) T(E) dE \, ,\nonumber 
\label{eq:PuLIV}
\end{eqnarray}
where the intensity normalized Stokes parameters $\Pq=\Pqdz$ and $\Pu=\Pudz$ depend on mass dimension $d$ and redshift $z_{\sourceindex}$ in the SME framework.

An upper bound on the observed polarization is then
\begin{eqnarray}
\pstar - 2 \sigma_{\star} < \Pmaxdz = \sqrt{ \Big(\Pqdz\Big)^2 + \Big(\Pudz\Big)^2 }\, ,
\label{eq:Pmaxdz}
\end{eqnarray}
such that observing a polarization fraction $\pstar$ implies an upper bound on $\kdV$ by finding the largest value of $\kdV$ that is consistent with the inequality $\Pmaxdz > \pstar - 2 \sigma_{\star}$, where $\sigma_{\star}$ is the 1-$\sigma$ uncertainty on the polarization measurement. This corresponds to a 95\% confidence interval assuming Gaussian measurement errors for the polarization fraction.

As shown by \cite{kislat17}, in this framework, broader filters lead to smaller values for $\Pmaxdz$, so observing larger $\pstar$ values in those filters leads to tighter constraints on $\kdV$ than observing the same polarization $\pstar$ through a narrower filter for the same source.  In addition to improving our signal-to-noise, this is a key reason we chose the broader $Lum$ band filter to compare to the more standard $I_c$ band filter, and implemented a method to combine both filters using simultaneous observations on two telescopes. The transmission $T(\lambda)$ for our combined $Lum+I_c$-band polarimetry is shown in Fig.~\ref{fig:Lum+I_TE}, which can be used to compute $T(E)$. Our observational setup is described in \S~\ref{sec:apol}.

In principle, one should also consider the source spectrum and the atmospheric attenuation in computing $T(E)$, but we follow \cite{kislat17} and assume that the optical spectra are flat enough in the relevant wavelength range so that we can ignore these small effects. However, unlike \cite{kislat17}, which only consider the transmission function of the broadband filter,
we additionally consider the transmission functions for the optics and CCD detector, in addition to the filter, when computing $T(E)$ (see Fig.~\ref{fig:apoltrans}). 

\begin{figure}
\centering
\begin{tabular}{@{}c@{}}

\includegraphics[width=3.5in]{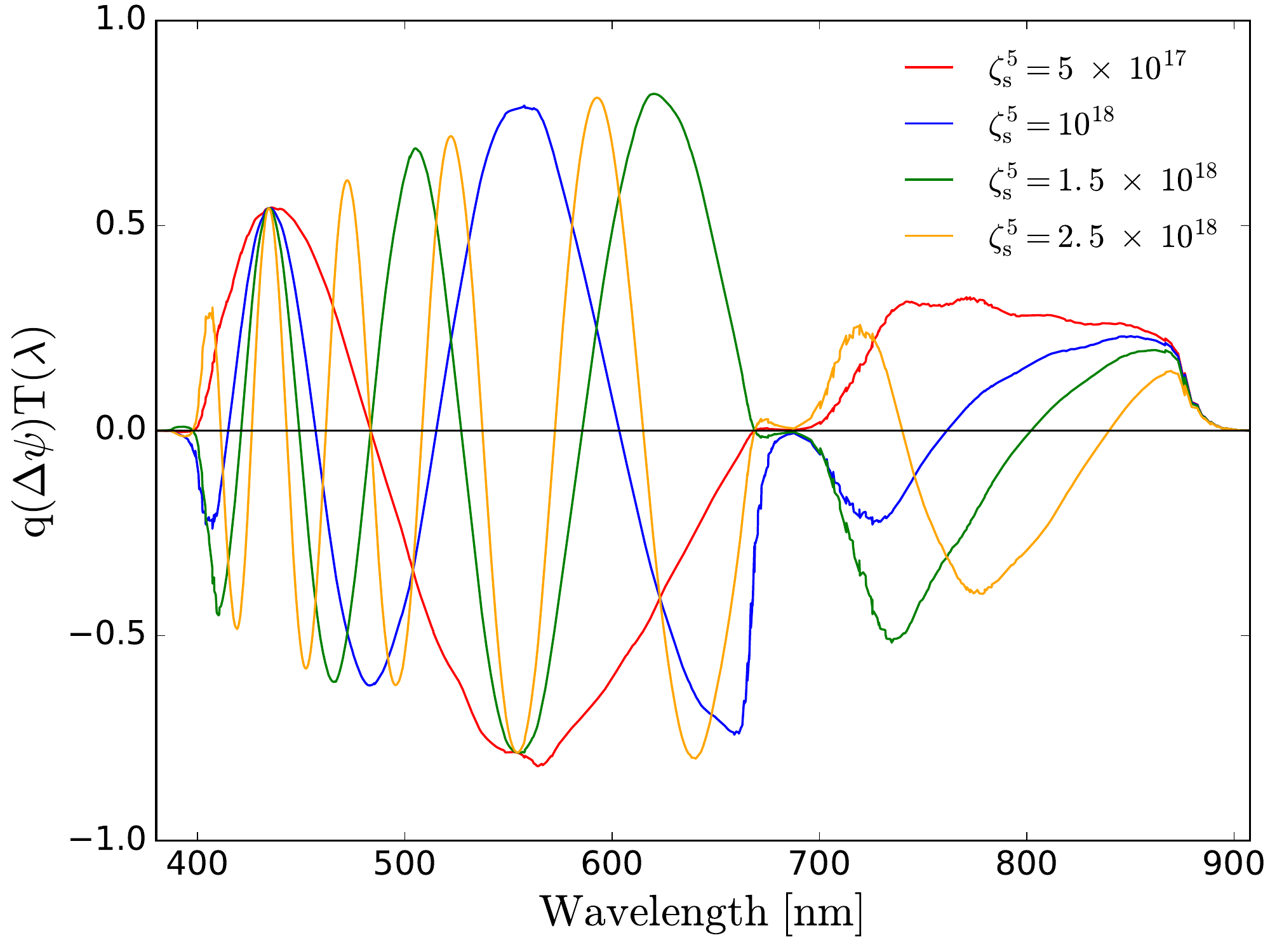} \\

\end{tabular}
\caption{
\baselineskip 9pt 
Change of the Stokes parameter $\Pqdz$ from Eq.~(\ref{eq:PqLIV}) for our combined $Lum+I_c$ filter in Fig.~\ref{fig:Lum+I_TE} for several values of $\zetakfive$. For comparison, see Fig. 2 of \cite{kislat17}.
}
\label{fig:DeltaStokes}
\end{figure}

\begin{figure}
\centering
\begin{tabular}{@{}c@{}}

\includegraphics[width=3.5in]{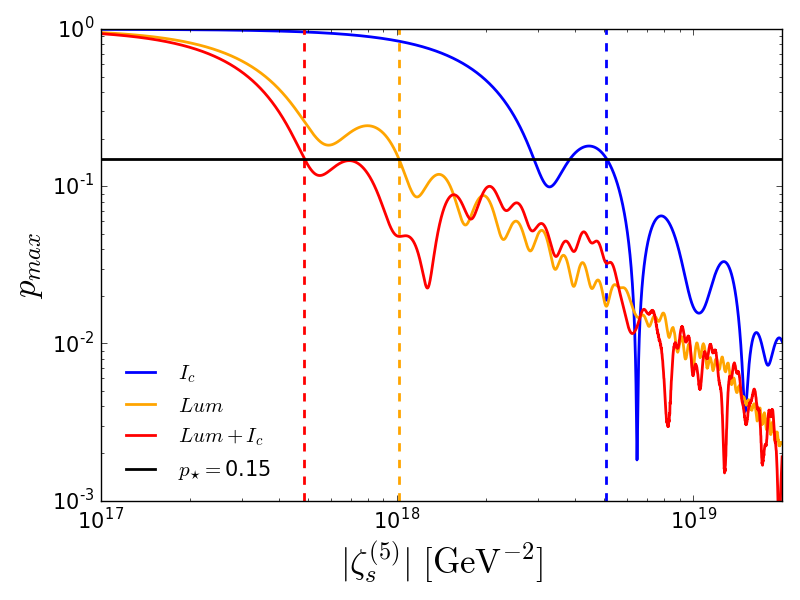} \\

\end{tabular}
\caption{
\baselineskip 9pt 
Maximum allowed polarization fraction $\pmax$ vs. $d=5$ CPT-Odd Vacuum birefringence parameter $|\zetakfive|$ from Eq.~(\ref{eq:zetak5}) for the $I_c$-band (blue), $Lum$-band (orange), and our combined $Lum+I_c$-band (red). For an 
example observed polarization fraction $\pstar=0.15$ (horizontal black line), upper limits on $|\zetakfive|$ for each band (dashed vertical lines) can be obtained by noting that $\pmax$ eventually falls below the observed value of $\pstar$ for all values of that coefficient. For $\pstar \gtrsim 0.02$, the most stringent upper limit comes from the combined $Lum+I_c$-band. For $\pstar=0.15$, this yields a $Lum+I_c$-band upper limit $|\zetakfive| \lesssim 5.0\times10^{17}$ GeV$^{-2}$, a factor of $\sim$$10$ better than the corresponding limit from the $I_c$ band.
}
\label{fig:d5maxpol}
\end{figure}

\begin{figure}
\centering
\begin{tabular}{@{}c@{}}

\includegraphics[width=3.5in]{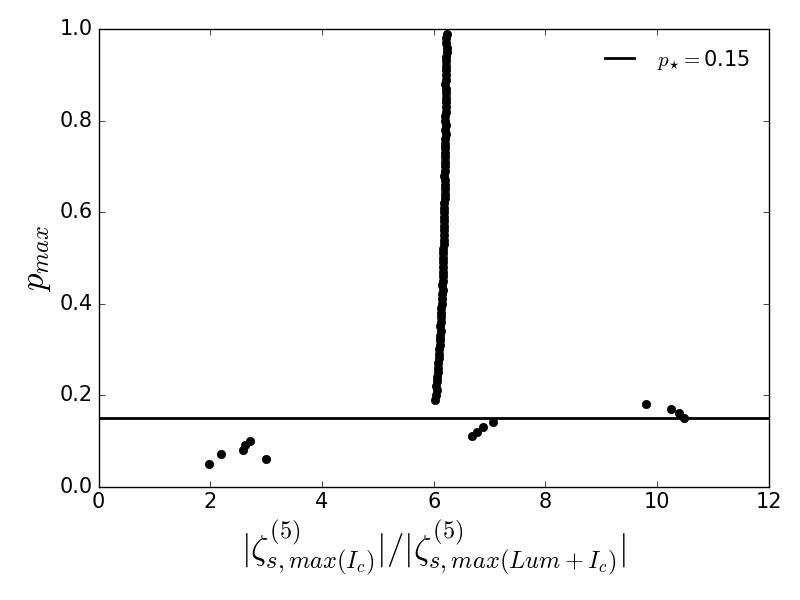} \\

\end{tabular}
\caption{
\baselineskip 9pt 
Theoretical maximum observed polarization $\pmax$ vs. the ratio of CPT-odd vacuum birefringent $d=5$ SME coefficients from Fig.~\ref{fig:d5maxpol} from the $I_c$ and $Lum+I_c$ bands, $|\zetakfive(I_c)|/|\zetakfive(Lum+I_c)|$. Ignoring polarization uncertainties $\sigma_{\star}$, for all observed polarization fractions $\pstar \gtrsim 0.02$ (where $\pstar < \pmax$), constraints from the $Lum+I_c$ band are $\sim$$2$-$10$ times tighter than for the $I_c$-band. The spike at $|\zetakfive(I_c)|/|\zetakfive(Lum+I_c)|\sim 6$ results from the fact that the ratio of the $|\zetakfive|$ values in each band (blue and red curves in Fig.~\ref{fig:d5maxpol}) is nearly constant for $\pstar \gtrsim 0.17$.
}
\label{fig:Lum+IcIcratiod5}
\end{figure}

Following \cite{kislat17}, to jointly parametrize the cosmological redshift dependence and SME parameter effects, we define the quantity $\zetakfive$ as
\begin{eqnarray}
\zetakfive \equiv \Ldzfive \kdVfive\, .
\label{eq:zetak5}
\end{eqnarray}
Also following \cite{kislat17}, Fig.~\ref{fig:DeltaStokes} shows the change in the intensity normalized Stokes parameter $\Pqdz$ from Eq.~(\ref{eq:PqLIV}) for several values of $\zetakfive$, while Fig.~\ref{fig:d5maxpol} shows theoretical limits from the maximum observed polarization $\pmax$ versus $\zetakfive$ in our $Lum$ and $I_c$ bands, and for our combined $Lum+I_c$-band in Fig.~\ref{fig:Lum+I_TE}. Based on Fig.~\ref{fig:d5maxpol}, Fig.~\ref{fig:Lum+IcIcratiod5} shows that the $Lum+I_c$ band yields $|\zetakfive|$ constraints $\sim$$2$-$10$ times more restrictive than the $I_c$ band for the same observed polarization fraction, over the range $\pmax \gtrsim 0.02$, (where $\pstar < \pmax$), assuming negligible uncertainties, $\sigma_{\star}$.

\subsection{CPT-Odd Vacuum Isotropic SME Models}
\label{sec:smevac}

Since $jm$ are the angular quantum numbers with $-j \le m \le j$, with $j \le d-2$, for each value of $d$, the number of distinct SME coefficients increases (See Table II of \cite{kostelecky13}). For example, the $d=5$ model has 16 SME coefficients \cite{kislat17}. Since we only observed two sources, we are limited to constraining only linear combinations of SME coefficients $\cdIjm$ and $\kdVjm$ along two specific lines of sight. Ultimately, progressively larger numbers of sources at different locations on the sky are required to better constrain the general anisotropic model space for a given value of $d$.

However, we can follow a simpler approach and also test the subset of isotropic models, which are recovered for each value of $d$ when setting $j=m=0$. Lines of sight to individual point sources are therefore most useful for constraining the isotropic SME coefficients $\cdIoo$ and $\kdVoo$, which correspond to some of the simplest LIV models in the literature \cite{kostelecky09,kostelecky13}. Constraints for both isotropic SME models and linear combinations along our specific lines of sight are shown in Table~\ref{tab:coefficients}.

\subsection{CPT-Even Vacuum Birefringent SME Models}
\label{sec:vacbiref2}

There exists an additional subset of CPT-even vacuum birefringent SME models with coefficients $\kdEjm$ and $\kdBjm$, where $d=4,6,8,\ldots$, which correspond to spin-2 helicity, rather than spin-0. Let us first define 
\begin{eqnarray}
\kdEB & \equiv & \Big| \sum_{jm} \twoYjmi \Big( \kdEjm + i\kdBjm \Big) \Big|\, ,
\label{eq:kdEB}
\end{eqnarray}
as shorthand for the absolute value of the linear combination of CPT-even birefringent SME coefficients for source $\sourceindex$. Note that there do not exist isotropic models for this subset of SME parameters so there are no $jm=00$ terms corresponding to Eq.~(\ref{eq:kdEB}). The CPT-even case is also more complex than the CPT-odd case, because the normal modes are linearly polarized and, in general, can involve no change in the polarization angle, or mixing of linearly polarized into elliptical or circularly polarized modes \cite{kostelecky13}.

We now define the accumulated phase change $\Phi$ at a given energy $E$ as
\begin{eqnarray}
\Phi_{\sourceindex} = 2 E^{d-3} \frac{\Ldz}{c} \kdEB\, .
\label{eq:PhiiEB}
\end{eqnarray}
In the CPT-odd vacuum birefringent case, this phase change directly resulted in a polarization angle rotation because we can split linearly polarized light equally into left and right circularly polarized states. But the same is not true of the CPT-even case. Linearly polarized light will not in general be split evenly between the normal modes of a CPT-even Lorentz violation.

But similar to the CPT-odd case, we can still arrive at an expression for the maximum allowed polarization given a particular broadband filter. Again assuming a 100\% polarized source at all wavelengths, the observation of a linear polarization fraction $\pstar$ (with uncertainty $\sigma_{\star}$) in a given broad energy band can be used to constrain the quantity $\kdEB$.

Following \cite{kostelecky13}, let us first define the angle $\Psi=\psi_0-\psi_b$ as the difference between the initial polarization angle $\psi_0$ for light not produced in a normal mode and the initial polarization angle $\psi_b$ for the slower of the two normal modes. For simplicity, we omit the source index $\sourceindex$ from the notation for $\Psi, \psi_0$, and $\psi_b$, since we will soon make assumptions which remove the $\Psi$ dependence.

Additionally we can define $\cosphiavg$ as the average value of $\cos \Phi_{\sourceindex}$ for source $\sourceindex$ after integrating over the relevant energy band
\begin{eqnarray}
\cosphiavg = \int_{E_1}^{E_2} \cos \Big( 2 \left(E^{d-3}{\scriptstyle -}E_1^{d-3}\right) \Ldz \kdEB \Big) T(E) dE \, . \nonumber \\
\label{eq:intcosphiEB}
\end{eqnarray}

\begin{figure}
\centering
\begin{tabular}{@{}c@{}}

\includegraphics[width=3.5in]{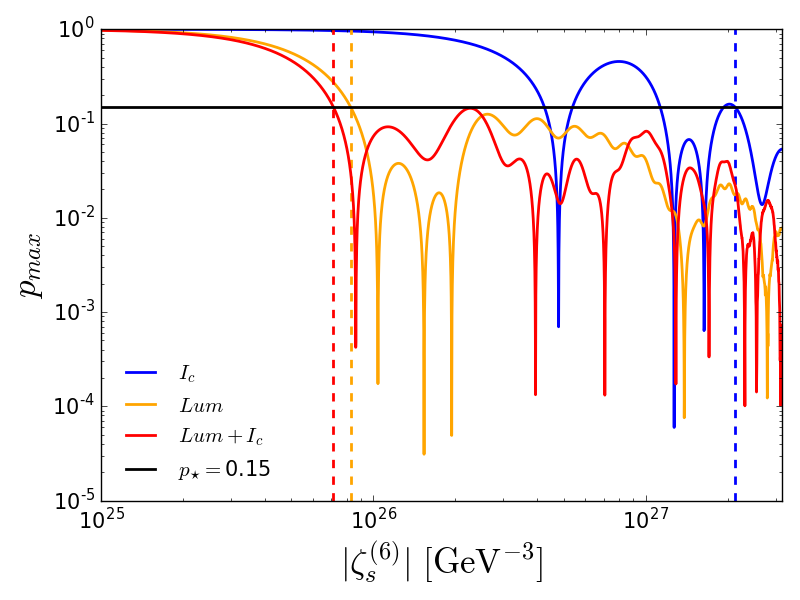} \\

\end{tabular}
\caption{
\baselineskip 9pt 
Similar to Fig.~\ref{fig:d5maxpol}, but for $\pmax$ vs. the $d=6$ CPT-Even Vacuum Birefringence parameter $|\zetaksix|$ from Eq.~(\ref{eq:zetak6}). For an example observed polarization fraction $\pstar=0.15$ (horizontal black line), the most stringent upper limit of $|\zetaksix| \lesssim 7\times10^{25}$ GeV$^{-3}$ (dashed red line) comes from our combined $Lum+I_c$-band, a factor of $\sim $$30$ better than the corresponding limit from the $I_c$ band (dashed blue line).
}
\label{fig:d6maxpol}
\end{figure}

In this case, as shown in Appendix~\ref{sec:david}, the normalized Stokes parameters $q=\Pqdz$ and $u=\Pudz$ are
\begin{eqnarray}
\Pqdz & = & \cos 2\Psi \cos 2\psi_b - \cosphiavg^2 \sin 2\Psi \sin 2\psi_b\, , \\
\Pudz & = &\cos 2\Psi \sin 2\psi_b + \cosphiavg^2 \sin 2\Psi \cos 2\psi_b\, ,
\label{eq:CPTevenQU}
\end{eqnarray}
and, via Eq.~(\ref{eq:Pmaxdz}), the corresponding maximum limit on polarization is
\begin{eqnarray}
& & \pstar - 2\sigma_{\star}  <  \Pmaxdz   \, \nonumber \\
& & = \sqrt{  1 - \left( 1 - \cosphiavg^2 \right) \sin^2 2\Psi  } \leq | \cosphiavg |\, ,  
\label{eq:maxpolEB}
\end{eqnarray}
where the conservative upper bound is reached when $\Psi=\pi/4$. Fig.~\ref{fig:d6maxpol} shows the corresponding limits obtained in this most conservative case.

\begin{figure}
\centering
\begin{tabular}{@{}c@{}}

\includegraphics[width=3.5in]{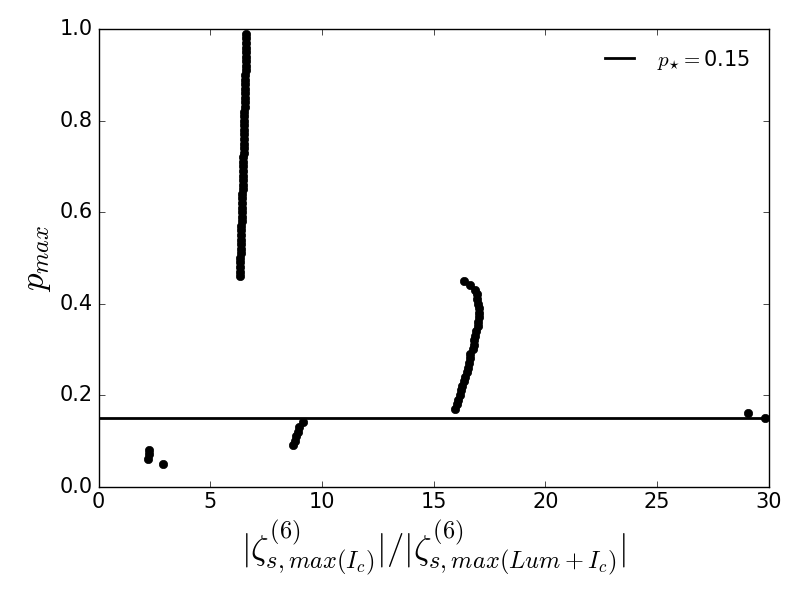} \\

\end{tabular}
\caption{
\baselineskip 9pt 
Similar to Fig.~\ref{fig:Lum+IcIcratiod5}, but for $\pmax$ vs. the ratio of CPT-even vacuum birefringent $d=6$ SME coefficients from Fig.~\ref{fig:d6maxpol} from the $I_c$ and $Lum+I_c$ bands, $|\zetaksix(I_c)|/|\zetaksix(Lum+I_c)|$. Again, ignoring polarization uncertainties $\sigma_{\star}$, for all observed polarization fractions $\pstar \gtrsim 0.1$ and for many values $\pstar < 0.1$ (where $\pstar < \pmax$), constraints from the $Lum+I_c$ band are $\sim$$3$-$30$ times tighter than for the $I_c$-band.
}
\label{fig:Lum+IcIcratiod6}
\end{figure}

Similar to Fig.~\ref{fig:d5maxpol}, Fig.~\ref{fig:d6maxpol} shows limits from the theoretical maximum polarization $\pmax$ in our $Lum$, $I_c$, and $Lum+I_c$-bands versus the quantity $\zetaksix$, defined as
\begin{eqnarray}
\zetaksix \equiv \Ldzsix \kdEBsix\, .
\label{eq:zetak6}
\end{eqnarray}
Again, similar to Fig.~\ref{fig:Lum+IcIcratiod5}, Fig.~\ref{fig:Lum+IcIcratiod6} shows that the combined $Lum+I_c$-band yields $|\zetaksix|$ constraints up to $\sim$$3$-$30$ times more sensitive than the $I_c$-band.

\subsection{Constraints on SME Models}
\label{sec:constraints}

With simultaneous photometric time series in two filter bands, one can estimate upper limits to any time delays (or early arrivals) between the corresponding light curves under the simple assumption that the intrinsic light curve shapes are identical. We perform this analysis on our entire photometric time series (see Tables~\ref{tab:data1}-\ref{tab:data2} and Figs.~\ref{fig:bllac1a}-\ref{fig:s5data1c}) using an open source implementation of the Discrete Correlation Function (DCF) in Python\footnote{\url{https://github.com/astronomerdamo/pydcf}}, which can be used to analyze variable time series with arbitrary sampling \cite{edelson88} (see, for example \cite{robertson15}). Constraints from time delays are presented in Table~\ref{tab:coefficients}, using the methods of \S\ref{sec:smevacdis}.

We consider possible estimated time delays $\Delta t_{\star} = m_I - m_L$ between observed photometric light curves in the $Lum$ and $I_c$ bands. Since our data points have a typical 8-10 minute cadence, we compute the best-fit DCF time delay using a series of DCF bin widths in the range $[5,20]$ minutes with step size $0.1$ minutes, while considering possible time delays or early arrivals in the range of $[-250,250]$ minutes for both sources. The mean and standard deviation of the set of best fit DCF time delays then yields $\Delta t_{\star}=$\tdelaybllac{} minutes and $\Delta t_{\star}=$\tdelaysfive{} minutes for BL Lacertae and \stwo{}, respectively. Both are consistent with $\Delta t_{\star}=0$, and thus no time delay, to within the $2$-$\sigma$ uncertainties. Using the $2$-$\sigma$ errors, and remaining agnostic as to the sign of $\Dtd$ leads to conservative time delay upper bounds of 
\begin{equation}
|\Delta t_{\star}| \le \max\{|\Delta t_{\star} - 2 \sigma_{\Delta t_{\star}}|,|\Delta t_{\star} + 2 \sigma_{\Delta t_{\star}}|\}\, , 
\label{eq:deltatup1}
\end{equation}
which for the two sources yields
$|\Dtd| \le |\Delta t_{\star}| = \tdelayupbllac$ minutes and $\tdelayupsfive$ minutes, for BL Lacertae and \stwo{}, respectively.\footnote{The time delay upper limit for BL Lacertae is less stringent than the limit form \stwo{} due mainly to the smaller number of data points.}

For a polarimetric time series measuring the polarization $p$ in either the $Lum$, $I_c$, or combined $Lum+I_c$-bands, one can use the maximum observed polarization $\pstar$ during the observing period to place limits on the SME parameters as in \S\ref{sec:smebiref}-\ref{sec:vacbiref2}, with an additional correction for systematic errors described in Sec.~\ref{sec:obs}. While a longer survey could, in principle, yield larger values of $\pstar$, and thus, more stringent SME constraints, meaningful constraints can still be obtained for arbitrary values of $\pstar$, even though these are likely lower limits to the true maximum polarization. Constraints from maximum observed polarization measurements are presented in Table~\ref{tab:coefficients} for the $Lum$ and $I_c$-bands, and in Table~\ref{tab:coefficients1} for the combined $Lum+I_c$-bands.

Even though we observed only two low redshift sources with small telescopes, our best $Lum$-band $d=5$ SME constraint from maximum polarization measurements of \stwo{} at $z=0.31 \pm 0.08$ in Tables~\ref{tab:coefficients}-\ref{tab:coefficients1}, of $\kdVfiveo < 1 \times 10^{-23}$ GeV$^{-1}$ is within an order of magnitude of all  constraints for individual lines of sight from the 36 QSOs in the redshift range $z\in[0.634,2.936]$ analyzed in Table II of \cite{kislat17}, where their SME parameter $\gamma_{\rm max}$ corresponds to our parameter $\kdVfiveo$.\footnote{Our $Lum+I_c$, $\kdVfiveo$ constraint is actually a factor of $\sim$$2$ worse than our $Lum$-band constraint because the maximum observed polarization for the combined $Lum+I_c$ band data of $\pstarcor =7.83 \pm 0.38 \%$ is slightly smaller than the $Lum$ band measurement of $\pstarcor = 9.77 \pm 0.52 \%$. For the same value of $\pstarcor$, the $Lum+Ic$ constraint will always be more sensitive than the $Lum$ or $I_c$-band constraints alone.} More specifically, our best $d=5$ constraint is comparable to the least sensitive constraint $\gamma_{\rm max} < 9.79 \times 10^{-24}$ GeV$^{-1}$ from Table II of \cite{kislat17} (for FIRST J21079-0620 with $\pstar = 1.12 \pm 0.22\%$ at $z=0.644$), while our best constraint is only a factor of $\sim$$10$ less sensitive than the {\it best} constraint of $\gamma_{\rm max} < 0.97 \times 10^{-24}$ GeV$^{-1}$ (for PKS 1256 − 229 with $\pstar=22.32 \pm 0.15 \%$ at $z=1.365$). This is the case even though our analysis was arguably more conservative than \cite{kislat17}, in regard to modeling our transmission functions, correcting for polarimetry systematics, and including uncertainties in the reported redshift measurements.

Note that the sources analyzed in \cite{kislat17} used linear polarization measurements from \cite{sluse05}, which were observed using the 3.6-m telescope at the European Southern Observatory in La Silla, with the EFOSC2 polarimeter equipped with a $V$-band filter.  As such, this work demonstrates that meaningful SME constraints for individual lines of sight --- that are comparable to, or within a factor of 10 as sensitive as polarimetry constraints from a $3.6$-m telescope --- can be readily obtained using a polarimetric $Lum+I_c$-band system of small telescopes with an effective $0.45$-m aperture, with $\sim$64 times less collecting area, which we describe in Sec.~\ref{sec:apol}.

\begin{table*}[h!]
\footnotesize
\begin{tabular}{ccccc} 
\hline
  \multicolumn{1}{c}{Source}
& \multicolumn{2}{c}{\stwo}
& \multicolumn{2}{c}{BL Lacertae}\\ 
  \multicolumn{1}{c}{$(RA,DEC)$}
& \multicolumn{2}{c}{($110.47^\circ$, $71.34^\circ$)}   
& \multicolumn{2}{c}{($330.68^\circ$, $42.28^\circ$)}\\
  \multicolumn{1}{c}{Redshift $z$}
& \multicolumn{2}{c}{0.31 $\pm$	0.08}   
& \multicolumn{2}{c}{0.0686	$\pm$ 0.0004}\\
\hline 
\hline 
\multicolumn{1}{c}{Time Delay Upper Bound} 
& \multicolumn{2}{c}{$Lum$-$I_c$}   
& \multicolumn{2}{c}{$Lum$-$I_c$}\\
$|\Delta t_{\star}|$ [minutes] & \multicolumn{2}{c}{\tdelayupsfive} & \multicolumn{2}{c}{\tdelayupbllac} \\
\hline
\hline
$\cdIsixo   \equiv| \sum_{jm} Y_{jm}(\theta, \phi) c_{(I)jm}^{(6)} |$ & \multicolumn{2}{c}{$< 6 \times 10^{+01}$ GeV$^{-2} $} & \multicolumn{2}{c}{$< 8 \times 10^{+02}$ GeV$^{-2} $} \\
$\cdIeighto \equiv| \sum_{jm} Y_{jm}(\theta, \phi) c_{(I)jm}^{(8)} |$ & \multicolumn{2}{c}{$< 8 \times 10^{+18}$ GeV$^{-4} $} & \multicolumn{2}{c}{$< 1 \times 10^{+20}$ GeV$^{-4} $} \\
$| c_{(I)00}^{(6)} |$ & \multicolumn{2}{c}{$< 2 \times 10^{+02}$ GeV$^{-2} $} & \multicolumn{2}{c}{$< 3 \times 10^{+03}$ GeV$^{-2} $} \\
$| c_{(I)00}^{(8)} |$ & \multicolumn{2}{c}{$< 3 \times 10^{+19}$ GeV$^{-4} $} & \multicolumn{2}{c}{$< 4 \times 10^{+20}$ GeV$^{-4} $} \\
\hline 
\hline 
  \multicolumn{1}{c}{Maximum Observed Polarization}
& \multicolumn{1}{c}{$Lum$}   
& \multicolumn{1}{c}{$I_c$}
& \multicolumn{1}{c}{$Lum$}   
& \multicolumn{1}{c}{$I_c$}\\
$\pstar$ [\%] & $10.02 \pm 0.44$ & $8.30 \pm 0.48$ & $10.33 \pm 0.43$ & $10.50 \pm 0.30$ \\
$\pisp$ [\%] & \pstartwoLumsfive & \pstartwoIcsfive & \pstartwoLumbllac & \pstartwoIcbllac \\
$p_{\rm sys,int}$ [\%] & 0.04 & 0.04 & 0.04 & 0.04 \\
\hline 
$\pstarcor$ [\%] & $9.77 \pm 0.52$ & $7.49 \pm 0.53$ & $9.37 \pm 0.44$ & $10.00 \pm 0.31$ \\
\hline 
\hline 
$\kdVfiveo  \equiv | \sum_{jm} Y_{jm}(\theta, \phi) k_{(V)jm}^{(5)} |$ & $< 1 \times 10^{-23}$ GeV$^{-1} $ & $< 7 \times 10^{-23}$ GeV$^{-1} $ & $< 3 \times 10^{-23}$ GeV$^{-1} $ & $< 1 \times 10^{-22}$ GeV$^{-1} $ \\
$\kdVseveno \equiv | \sum_{jm} Y_{jm}(\theta, \phi) k_{(V)jm}^{(7)} |$ & $< 2 \times 10^{-6}$ GeV$^{-3} $ & $< 1 \times 10^{-5}$ GeV$^{-3} $ & $< 4 \times 10^{-6}$ GeV$^{-3} $ & $< 2 \times 10^{-5}$ GeV$^{-3} $ \\
$\kdVnineo  \equiv | \sum_{jm} Y_{jm}(\theta, \phi) k_{(V)jm}^{(9)} |$ & $< 3 \times 10^{+11}$ GeV$^{-5} $ & $< 4 \times 10^{+12}$ GeV$^{-5} $ & $< 8 \times 10^{+11}$ GeV$^{-5} $ & $< 6 \times 10^{+12}$ GeV$^{-5} $ \\
$| k_{(V)00}^{(5)} |$ & $< 5 \times 10^{-23}$ GeV$^{-1} $ & $< 3 \times 10^{-22}$ GeV$^{-1} $ & $< 1 \times 10^{-22}$ GeV$^{-1} $ & $< 4 \times 10^{-22}$ GeV$^{-1} $ \\
$| k_{(V)00}^{(7)} |$ & $< 6 \times 10^{-6}$ GeV$^{-3} $ & $< 3 \times 10^{-5}$ GeV$^{-3} $ & $< 2 \times 10^{-5}$ GeV$^{-3} $ & $< 8 \times 10^{-5}$ GeV$^{-3} $ \\
$| k_{(V)00}^{(9)} |$ & $< 1 \times 10^{+12}$ GeV$^{-5} $ & $< 1 \times 10^{+13}$ GeV$^{-5} $ & $< 3 \times 10^{+12}$ GeV$^{-5} $ & $< 2 \times 10^{+13}$ GeV$^{-5} $ \\
$\kdEBfouro  \equiv | \sum_{jm} {}_2 Y_{jm}(\theta, \phi) (k_{(E)jm}^{(4)} + ik_{(B)jm}^{(4)}) |$ & $\lesssim 7 \times 10^{-32} $ & $\lesssim 2 \times 10^{-31} $ & $\lesssim 2 \times 10^{-31} $ & $\lesssim 3 \times 10^{-31} $ \\
$\kdEBsixo   \equiv | \sum_{jm} {}_2 Y_{jm}(\theta, \phi) (k_{(E)jm}^{(6)} + ik_{(B)jm}^{(6)}) |$ & $\lesssim 5 \times 10^{-15}$ GeV$^{-2} $ & $\lesssim 2 \times 10^{-14}$ GeV$^{-2} $ & $\lesssim 1 \times 10^{-14}$ GeV$^{-2} $ & $\lesssim 5 \times 10^{-14}$ GeV$^{-2} $ \\
$\kdEBeighto \equiv | \sum_{jm} {}_2 Y_{jm}(\theta, \phi) (k_{(E)jm}^{(8)} + ik_{(B)jm}^{(8)}) |$ & $\lesssim 2 \times 10^{+2}$ GeV$^{-4} $ & $\lesssim 5 \times 10^{+3}$ GeV$^{-4} $ & $\lesssim 4 \times 10^{+2}$ GeV$^{-4} $ & $\lesssim 1 \times 10^{+4}$ GeV$^{-4} $ \\
\hline 
\end{tabular} 
\caption{
\footnotesize
Upper limits on linear combinations of SME coefficients of Lorentz and CPT violation along specific lines of sight from our $Lum$ and $I_c$-band observations of BL Lacertae and \stwo, with sky coordinates and redshifts from Table~\ref{tab:params}. The upper portion of the table shows the vacuum dispersion coefficients $\cdIo$ and corresponding isotropic coefficients $\cdIoo$ (see Fig.~\ref{fig:opticalbandSME}) as inferred from estimates of an upper bound on the time delay $\Delta t_{\star}$ between the observed photometry in both the $Lum$ and $I_c$-bands as described in Sec.~\ref{sec:constraints}. The remaining rows show separate constraints from the maximum observed polarization fraction $\pstar$, which we correct for systematics from interstellar polarization ($\pisp$) and instrumental polarization and zero point bias ($\pinst$), in each band via $\pstarcor = \pstar - \pisp - \pinst$, with corresponding statistical errors added in quadrature. To be conservative, we derive SME parameter upper bounds using the 2-$\sigma$ errors for $\Delta t_{\star}$, $\pstarcor$, and Redshift $z$. The lower rows show the vacuum birefringent coefficients $\kdVo$ and their corresponding isotropic coefficients $\kdVoo$, each for the CPT-odd cases $d=5,7,9$. The last three rows show the vacuum birefringence coefficients $\kdEBo$ for the CPT-even cases $d=4,6,8$. In each case, constraints from our observed broadband polarimetry using the wider $Lum$-band are tighter than for the $I_c$-band.
}
\label{tab:coefficients} 
\end{table*} 
\renewcommand{\arraystretch}{1} 

\begin{table*} [h!]
\footnotesize
\begin{tabular}{ccc} 
\hline
  \multicolumn{1}{c}{Source}
& \multicolumn{1}{c}{\stwo}
& \multicolumn{1}{c}{BL Lacertae}\\ 
  \multicolumn{1}{c}{$(RA,DEC)$}
& \multicolumn{1}{c}{($110.47^\circ$, $71.34^\circ$)}   
& \multicolumn{1}{c}{($330.68^\circ$, $42.28^\circ$)}\\
  \multicolumn{1}{c}{Redshift $z$}
& \multicolumn{1}{c}{0.31 $\pm$	0.08}   
& \multicolumn{1}{c}{0.0686	$\pm$ 0.0004}\\
\hline 
\hline 
  \multicolumn{1}{c}{Maximum Observed Polarization}
& \multicolumn{1}{c}{$Lum+I_c$}   
& \multicolumn{1}{c}{$Lum+I_c$}\\
$\pstar$ [\%] & $8.64 \pm 0.30$  & $10.30 \pm 0.28$  \\
$\pisp$ [\%] & \pstartwoIcsfive  & \pstartwoLumbllac  \\
$\pinst$ [\%] & 0.04  & 0.04 \\
\hline
$\pstarcor$ [\%] & $7.83 \pm 0.38$  & $9.34 \pm 0.29$  \\
\hline 
\hline 
$\kdVfiveo  \equiv | \sum_{jm} Y_{jm}(\theta, \phi) k_{(V)jm}^{(5)} |$ & $< 2 \times 10^{-23}$ GeV$^{-1} $ & $< 5 \times 10^{-23}$ GeV$^{-1} $ \\
$\kdVseveno \equiv | \sum_{jm} Y_{jm}(\theta, \phi) k_{(V)jm}^{(7)} |$ & $< 4 \times 10^{-6}$ GeV$^{-3} $ & $< 8 \times 10^{-6}$ GeV$^{-3} $ \\
$\kdVnineo  \equiv | \sum_{jm} Y_{jm}(\theta, \phi) k_{(V)jm}^{(9)} |$ & $< 8 \times 10^{+11}$ GeV$^{-5} $ & $< 2 \times 10^{+12}$ GeV$^{-5} $ \\
$| k_{(V)00}^{(5)} |$ & $< 9 \times 10^{-23}$ GeV$^{-1} $ & $< 2 \times 10^{-22}$ GeV$^{-1} $ \\
$| k_{(V)00}^{(7)} |$ & $< 1 \times 10^{-5}$ GeV$^{-3} $ & $< 3 \times 10^{-5}$ GeV$^{-3} $ \\
$| k_{(V)00}^{(9)} |$ & $< 3 \times 10^{+12}$ GeV$^{-5} $ & $< 7 \times 10^{+12}$ GeV$^{-5} $ \\
$\kdEBfouro  \equiv | \sum_{jm} {}_2 Y_{jm}(\theta, \phi) (k_{(E)jm}^{(4)} + ik_{(B)jm}^{(4)}) |$ & $\lesssim 8 \times 10^{-32} $ & $\lesssim 2 \times 10^{-31} $ \\
$\kdEBsixo   \equiv | \sum_{jm} {}_2 Y_{jm}(\theta, \phi) (k_{(E)jm}^{(6)} + ik_{(B)jm}^{(6)}) |$ & $\lesssim 9 \times 10^{-15}$ GeV$^{-2} $ & $\lesssim 5 \times 10^{-15}$ GeV$^{-2} $ \\
$\kdEBeighto \equiv | \sum_{jm} {}_2 Y_{jm}(\theta, \phi) (k_{(E)jm}^{(8)} + ik_{(B)jm}^{(8)}) |$ & $\lesssim 2 \times 10^{+2}$ GeV$^{-4} $ & $\lesssim 4 \times 10^{+2}$ GeV$^{-4} $ \\
\hline 
\end{tabular}
\caption{
\footnotesize
Same as SME coefficient limits from maximum observed polarization from Fig.~\ref{tab:coefficients1}, but for the combined $Lum$+$I_c$-band (see Fig.~\ref{fig:Lum+I_TE}).
}
\label{tab:coefficients1} 
\end{table*} 
\renewcommand{\arraystretch}{1} 

\section{The Array Photo Polarimeter}
\label{sec:apol}

The observing system used in this work, the Array Photo Polarimeter (\apol) --- maintained and operated by one of us (G. Cole) --- uses dual beam inversion optical polarimetry with Savart plate analyzers rotated through an image sequence with various half-wave-plate (HWP) positions. See \cite{tinbergen05,berry05,berry14} for the basic procedures underlying dual beam polarimetry. This approach can be contrasted with quadruple beam analyzers with Wollaston prisms such as RoboPol (e.g.~\cite{king14,panopoulou15,skalidis18}) that can obtain all the Stokes parameters in a suitably calibrated single image.

The \apol{} array employs an automated telescope, filter, and instrument control system with 5 co-located telescopes on two mounts. \apol{} uses two small, Celestron 11 and 14 inch, primary telescopes (C11 and C14) for polarimetry with an effective collecting area equivalent to a larger 17.8 inch (0.45-m) telescope, with added capability to obtain simultaneous photometry or polarimetry on a third smaller telescope (Celestron 8 inch = C8), along with bright star photometry and/or guiding using a fourth and fifth 5 inch telescope. \apol{} is located at StarPhysics Observatory (Reno, Nevada) at an elevation of 1585 meters. 

Earlier iterations of \apol{} (e.g.~\cite{cole10}) have been progressively equipped with new automated instrumentation and image reduction software \cite{cole07,cole08,cole09}, and used for spectropolarimetry studies \cite{cole01,cole16}, including a long observing campaign presenting polarimetry and photometry of the variable star Epsilon Aurigae \cite{cole11,cole12,cole13}. \apol{}'s polarimeter designs also helped inform the planning and hardware implementation of the University of Denver DUSTPol instrument, an optical polarimeter with low instrumental polarization that has been used to study cool star systems, including RS CVn systems and Wolf-Rayet stars \cite{wolfe15}.

\begin{figure*}
\centering
\begin{tabular}{@{}c@{}c@{}}

\includegraphics[width=3.5in]{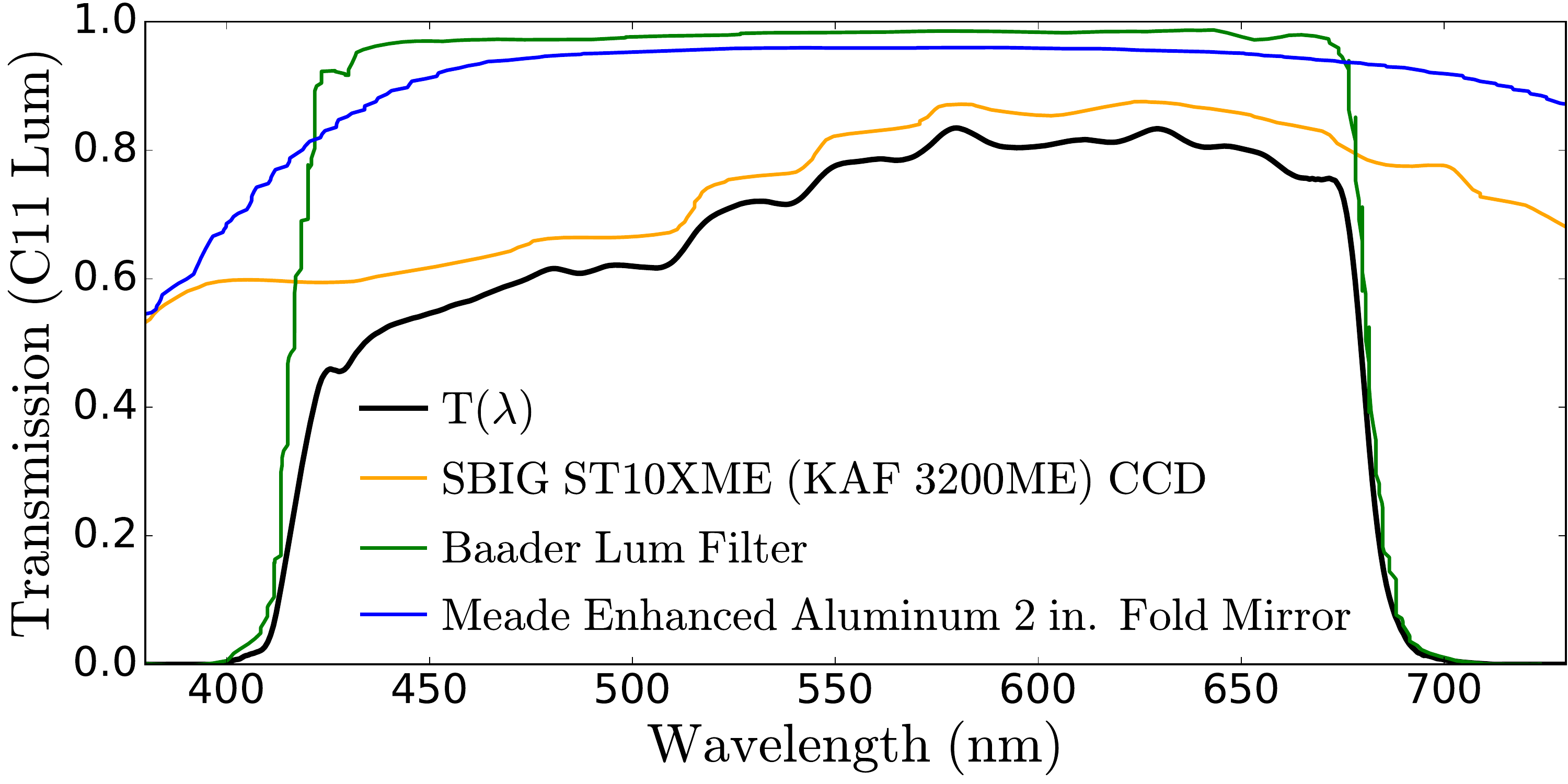} &
\includegraphics[width=3.5in]{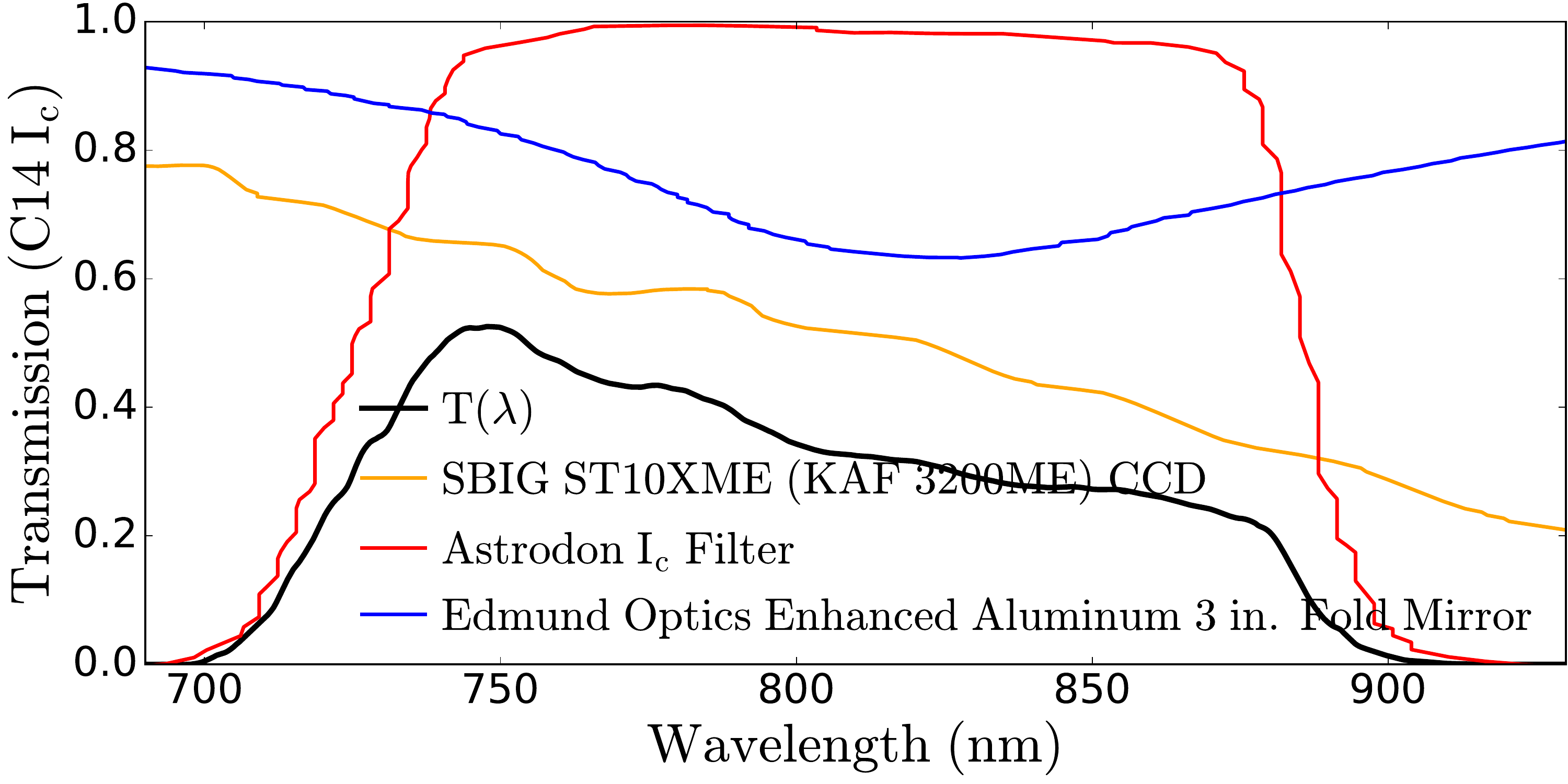} \\

\includegraphics[width=3.5in]{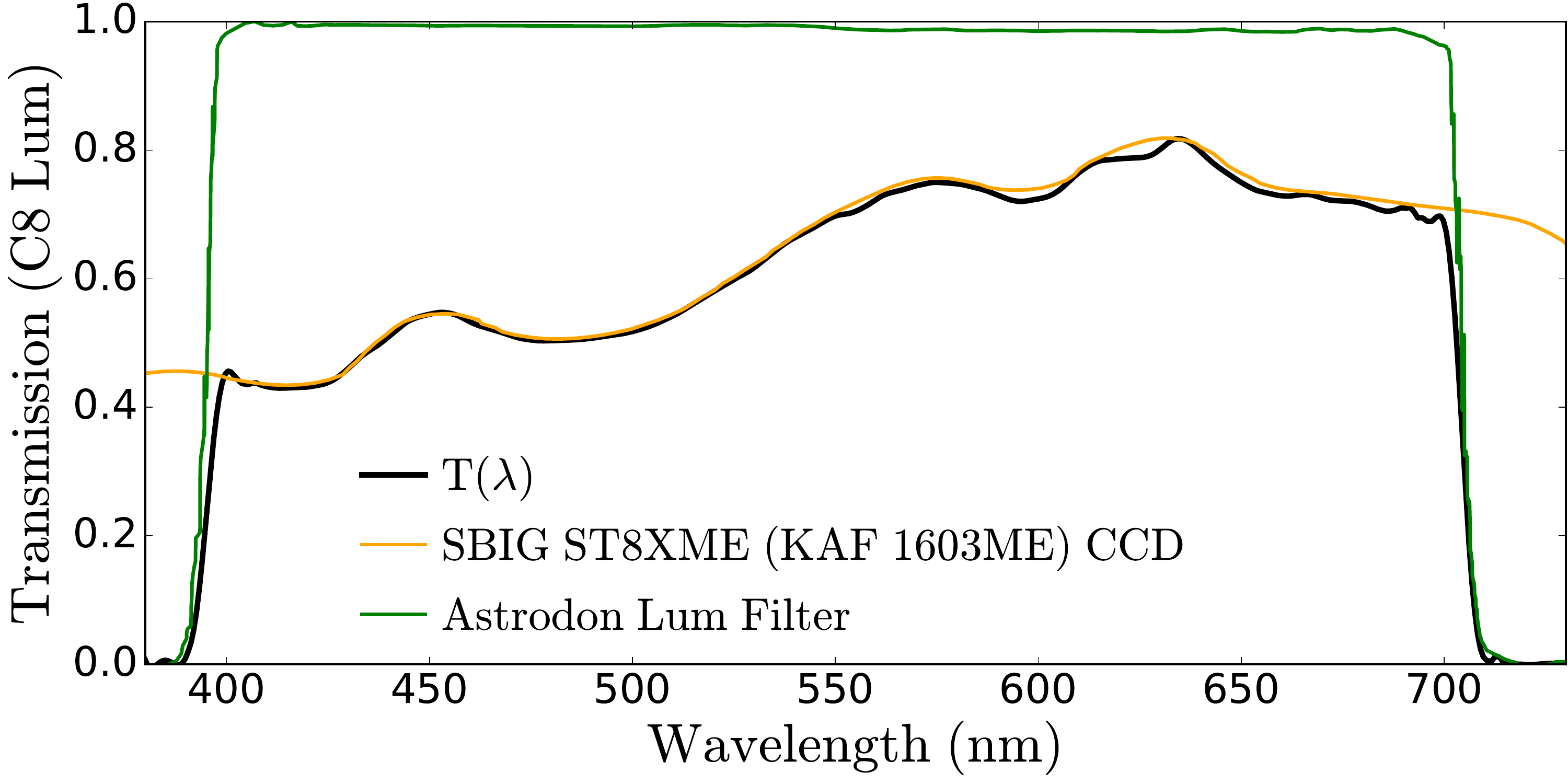} &
\includegraphics[width=3.5in]{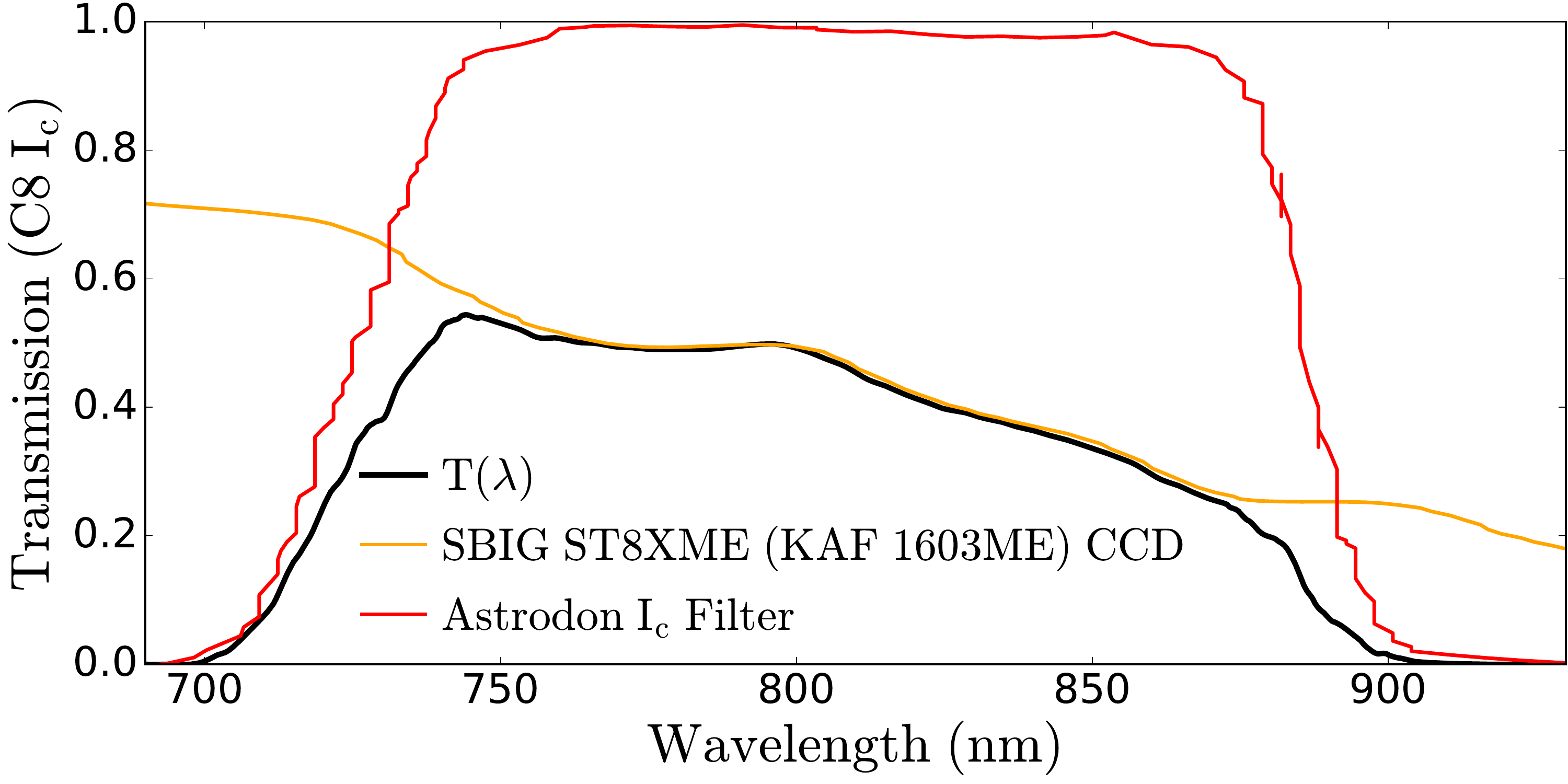} \\

\end{tabular}
\caption{
\baselineskip 9pt 
Array Photo Polarimeter transmission functions for the relevant optical components, filters, and CCD detectors. The first row shows the C11 and C14 telescopes used for polarimetry in the $Lum$ and $I_c$ bands, respectively. The second row shows the C8 telescope used for photometry in both the $I_c$ and $Lum$ bands. The black curve shows $T(\lambda)$, the total transmission function, which can be used to compute $T(E)$ to constrain the SME parameters via maximum polarization measurements as in \S~\ref{sec:smebiref}-\ref{sec:constraints}. To simplify the analysis, we do not model the transmission functions of the Celestron StarBright coatings. We also do not model the transmission functions of the Savart Plates and the half-wave plates, which are fairly uniform throughout the relevant wavelength range. For similar reasons, we also neglect the atmospheric transmission or the source spectra. By comparison, the analysis in \cite{kislat17} only included the filter transmission function.
}
\label{fig:apoltrans}
\end{figure*}

The first row of Fig.~\ref{fig:apoltrans} shows the inputs to the total transmission vs. wavelength $T(\lambda)$ in Fig.~\ref{fig:Lum+I_TE} for our $Lum$ and $I_c$-band polarimetry using \apol, which can be used to compute $T(E)$ as used in \S~\ref{sec:smebiref}-\ref{sec:constraints}. The \apol{} setup used in this work and the associated polarimetry data reduction and analysis methods will be described in more detail in a companion paper \cite{friedman18b}.

\subsection{Observations and Systematics}
\label{sec:obs}

All data in this paper were observed with \apol{} over a short campaign in December 2017 - January 2018. Samples of the observed data for BL Lacertae (3 nights: December 14, 15, 18 (2017)) and \stwo{} (5 nights: December 12, 13, 14, 15 (2017), January 1 (2018)) are shown in Appendix~\ref{sec:data1} in Tables~\ref{tab:data1}-\ref{tab:data2}, with the full machine-readable data available online and plotted in Figs.~\ref{fig:bllac1a}-\ref{fig:bllac1c} for BL Lacertae and in Figs.~\ref{fig:s5data1a}-\ref{fig:s5data1c} for \stwo{}. Image sequences with detected cosmic rays were identified as outliers and excluded. While we only use the maximum observed polarization to constrain birefringent SME models, we include the entire time series for completeness. By contrast, the full photometric time series was used to constrain the vacuum dispersion SME models using estimated time delays.

Optical photometric and polarimetric variability, correlations between flux and color, and searches for intra-band photometric time lags, have been studied extensively in the literature for AGN and BL Lacertae type objects \cite{sluse05,uemura10,falomo14,hovatta16,kokubo17,carnerero17}, including the specific, well known AGN sources we observed: BL Lacertae \cite{moore82,ikejiri11,zhang13,zhang16}, and \stwo{} \cite{impey00,nesci02,sasada08,larionov13,chandra15,bhatta15,bhatta16,doroshenko17,yuan17}.
Our analysis is restricted to testing SME models, but our photometric and polarimetric time series could be analyzed similarly in future work.\footnote{For reviews of the many other applications of optical polarimetry, see for example, \cite{hough05,hough06,hough07,hough11,bagnulo09,canovas11}.} 

Our data reduction pipeline removes systematic instrumental polarization using secondary flat-field self-calibration from the two sets of images taken at the 4 half wave plate positions ($0^{\circ}$, $22.5^{\circ}$) and ($45^{\circ}$, $67.5^{\circ}$), respectively, following \cite{berry14}. Hundreds of previous \apol{} measurements of unpolarized standard stars indicate that this procedure yields instrumental polarization systematics $\lesssim$ 0.03\% for targets with sufficient flux, while zero-point bias adjustments are typically $\lesssim$ 0.01\% for observed \apol{} polarization fractions of greater than a few percent \cite{cole12}.\footnote{We assume the same systematic error budget of 0.04\% for the $Lum$, $I_c$, and $Lum+I_c$ bands including instrumental polarization and zero-point bias.} The \apol{} HWP waveplate modulation efficiencies have been measured to be $\gtrsim 97\%$ and $\sim 90\%$ for $Lum$ and $I_c$, respectively. Since imperfect modulation efficiency can only reduce the maximum observed polarization from its true value, to be conservative, we choose not to model these systematics here.\footnote{HWP modulation efficiency systematics would not effect the measured polarization angles, although other relevant systematics are discussed in \cite{cole12}.} Previous tests indicate that other potential systematics including coordinate frame misalignment are negligible for \apol{} \cite{cole12}.

The total optical polarization along arbitrary lines of sight toward galactic field stars can range from a fraction of a percent to several percent \cite{weitenbeck08,meade12,siebenmorgen18}. Previous work from the Large Interstellar Polarization Survey provided evidence that interstellar polarization (ISP) from multiple dust clouds along a given line of sight is smaller than from lines of sight passing through a single dust cloud  \cite{bagnulo17,siebenmorgen18}. Since the presence of two or more clouds would therefore depolarize the incoming radiation, we assume that the ISP along the line of sight toward a galactic field star represents a conservative upper limit to the true ISP toward an AGN source that would have been measured through the full dust column of the galaxy, the intergalactic medium, and the AGN host galaxy.

Using a sample image sequence for each of our two AGN targets, we performed $Lum$ and $I_c$-band polarimetry on the two closest field stars within 3 arcmin of the target AGN, finding polarizations of \pstaroneLumbllac{}\% and \pstartwoLumbllac{}\% in $Lum$ and \pstaroneIcbllac{}\% and \pstartwoIcbllac{}\% in $I_c$ for the field of BL Lacertae 
and \pstaroneLumsfive{}\% and \pstartwoLumsfive{}\% in $Lum$ and \pstaroneIcsfive{}\% and \pstartwoIcsfive{}\% in $I_c$ for the field of \stwo{}. See Table~\ref{tab:comps}. For the combined $Lum+I_c$ band maximum polarization measurements, we use the largest ISP systematic from the $Lum$ and $I_c$ bands.

\begin{table*}[t]
\centering 
\begin{tabular}{cccccc} 
\hline \hline
  \multicolumn{1}{c}{Star \#}    
& \multicolumn{1}{c}{GAIA DR2 ID}  
& \multicolumn{1}{c}{RA}    
& \multicolumn{1}{c}{DEC} 
& \multicolumn{1}{c}{$p_L$} 
& \multicolumn{1}{c}{$p_I$} \\ 
  \multicolumn{1}{c}{} 
& \multicolumn{1}{c}{} 
& \multicolumn{1}{c}{IRCS(J2000)$^{\circ}$}   
& \multicolumn{1}{c}{IRCS(J2000)$^{\circ}$}    
& \multicolumn{1}{c}{(\%)} 
& \multicolumn{1}{c}{(\%)} \\ 
\hline 
\multicolumn{6}{c}{BL Lacertae} \\
\hline
1 &	1960066324769508992 & 330.68924090  &	+42.27652024   &    \pstaroneLumbllac &  \pstaroneIcbllac		\\
2 &	1960066329068001536 & 330.69304715  &	+42.28231354   &    \pstartwoLumbllac &  \pstartwoIcbllac \\		
\hline 
\multicolumn{6}{c}{\stwo{}} \\
\hline
1 &	1111278261916148224 & 110.47651216  &	+71.32247695   &    \pstaroneLumsfive &  \pstaroneIcsfive		\\
2 &	1111278158836933888 & 110.46803636  &	+71.30492029   &    \pstartwoLumsfive &  \pstartwoIcsfive \\	
\hline 
\hline 
\end{tabular} 
\caption{
Polarization measurements $p_L$ ($Lum$) and $p_I$ ($I_c$) of field stars in sample image sequences within 3 arcmin of the AGN sources BL Lacertae and \stwo, used to estimate upper limits on interstellar polarization for the systematic error budget used in Tables~\ref{tab:coefficients}-\ref{tab:coefficients1}. Celestial coordinates and GAIA DR2 identifiers from Simbad/VizieR are included.
}
\label{tab:comps} 
\end{table*} 

Assuming that our AGN max polarization measurements arise from a combination of instrumental polarization, zero point bias, ISP, and intrinsic source polarization, and that the ISP is approximately constant within 3 arcmin of the AGN target line of sight, we use the smaller of the two measured stellar polarizations to estimate conservative systematic upper limits for ISP ($\pisp$) as listed in Tables~\ref{tab:coefficients}-\ref{tab:coefficients1}. Finally, to obtain a polarization estimate corrected for systematics, $\pstarcor$, we subtract these systematic error estimates for ISP ($\pisp$) as well as the 0.04\% systematic budget from instrumental polarization and zero point bias ($\pinst$) from our maximum observed polarization in the $I_c$, $Lum$, and $Lum+I_c$-bands to obtain the SME constraints in Tables~\ref{tab:coefficients}-\ref{tab:coefficients1}.\footnote{Statistical errors from the polarization measurements for the AGN source and stars used to estimate ISP systematics are added in quadrature.}

\section{Conclusions}
\label{sec:conc}
In this work, we performed optical polarimetry and photometry of two well known AGN sources, BL Lacertae and \stwo{} in both the $Lum$ and $I_c$-bands, while implementing a procedure to obtain polarimetry in a wider effective passband with coverage from $\sim$$400$-$900$\,nm by combining simultaneous photometry from two small, co-located telescopes. We used the "average polarization" method of \cite{kislat17}, which analyzed polarimetric measurements from the literature, to analyze our own polarimetric measurements, thereby demonstrating a proof-of-principle method to use our own data to derive meaningful constraints, for individual lines of sight or isotropic models, on parameters from various subsets of the Standard Model Extension, a useful framework to test for new physics beyond the Standard Model including potential violation of Lorentz and CPT Invariance \cite{kostelecky09}. We demonstrated that maximum polarization measurements with our wider effective $Lum+I_c$ bandpass can yield SME constraints that are up to $\sim$$10$ or $\sim$$30$ times more sensitive that with our $I_c$-band filter, for $d=5$ and $d=6$ models, respectively.

To constrain SME parameters for a single source along a single line of sight, optical photometric measurements of AGN are not competitive with GRB gamma-ray and x-ray measurements in regard to timing resolution, energy, and redshift. Therefore, high energy GRB measurements are the best way to constrain SME parameters using observed time delays at different observed energies. Nevertheless, GRBs are transients both in their prompt gamma-ray emission and optical afterglows. Therefore, since AGN are the brightest continuous optical sources at cosmological distances, it is considerably easier to quickly obtain more complete sky coverage by observing many more AGN, in order to better constrain the anisotropic vacuum dispersion SME models. In addition, compared with gamma and x-ray polarimetry, optical polarimetric measurements typically have smaller statistical uncertainties and independent systematics \cite{kislat17}. Optical polarimetry is also easier to obtain with ground based instruments than gamma-ray and x-ray polarimetry, which must be obtained from space (e.g. \cite{toma12}). 

Although the limits presented here were not intended to compete with other approaches using maximum polarization measurements integrated over an optical bandpass, the pilot program in this work nevertheless demonstrates that meaningful SME constraints can be obtained even with a small set of telescopes with an effective 0.45-m aperture, that are competitive --- to within a factor of $\sim$$1$-$10$ in sensitivity for $d=5$ models --- even when compared to optical polarimetry from a $3.6$-m telescope \cite{sluse05,kislat17}. Since $d=6$ models were not analyzed in \cite{kislat17}, it would be interesting to perform similar comparisons to our $d=6$ constraints in future work. As such, there is a strong science case to use the maximum observed polarization for a large sample of AGN with wide optical bandpasses to constrain the anisotropic vacuum birefringent SME models, which include the three families of SME coefficients not constrained by time delay estimates. 

Future work could improve upon existing SME constraints simply by using the methods in this work to analyze optical polarimetry from large published surveys of AGN and quasars (e.g  \cite{angelakis16,hutsemekers17}) in addition to the AGN and GRB afterglow sources already studied by \cite{kislat17}. In addition, state-of-the-art SME constraints could potentially be obtained by performing a new survey to significantly increase the number of high redshift sources with published optical polarimetry along independent lines of sight. The pilot program described in this work thus serves to motivate a dedicated optical AGN polarimetric survey similar to the Steward Observatory spectrapolarimetric AGN monitoring program \cite{sluse05}, the RoboPol survey of gamma-ray selected blazars \cite{king14,blinov15,angelakis16}, or the La Silla Observatory survey of optical linear polarization of QSOs \cite{sluse05,hutsemekers17}, to name some relevant examples.

Such future surveys would obtain broadband optical polarimetry of each AGN source with a set of filters, optics, and detectors optimally chosen to improve upon the SME constraints obtainable using the more standard optical filters employed by previous surveys. In addition to measuring sources along lines of sight without previously published polarimetry, where possible, polarimetric measurements of previously observed sources could still lead to tighter SME constraints by either observing a larger maximum polarization than what was reported in the literature, or by observing  with a wider optical bandpass.

By duplicating this setup on one or more 1-meter class telescopes in each hemisphere, using the same data reduction software, such a survey could achieve the full sky coverage needed to fully constrain the more general anisotropic SME models at increasingly larger mass dimension $d \geq 5$. However, unlike previous surveys, it may only be necessary to observe a short duration time series for each AGN source, in order to maximize the number of sources with maximum polarization measurements, thereby optimizing a to-be-determined figure of merit which would quantify the improvement in constraints for specific SME models, during a given survey time period. 

Since spectropolarimetry typically yields SME $d=5$ model parameter constraints that are $\sim$$2$-$3$ orders of magnitude more sensitive than using a single, broadband, optical filter \cite{kislat17}, it would also be interesting to investigate the costs and benefits of a full spectropolarimetric survey on $\gtrsim 2$-m class telescopes versus a less expensive, shorter duration, survey on a set of 1-m class telescopes using multiple optical filters to test $d \ge 5$ SME models. Similarly, it would be worthwhile in future work to explore the tradeoffs for constraining SME models by using multiple, non-overlapping, narrow-band, optical filters to effectively perform low resolution spectropolarimetry versus combining two or more filters into a single, broadband filter, as demonstrated in this work.

Design feasibility studies for such a proposed survey will be analyzed in future work, with emphasis on the best path to quickly achieve the largest payoff for astrophysical tests of CPT and Lorentz Invariance violation without the time and expense required to perform an all sky spectrapolarimetric survey.
\acknowledgments

This work was originally inspired in part by conversations with Chris Stubbs. We thank Calvin Leung for sharing code to help compute transmission functions. A.S.F. acknowledges support from NSF INSPIRE Award PHYS 1541160. B.G.K. acknowledges support from UCSD's Ax Center for Experimental Cosmology. We gratefully made use of the NASA/IPAC Extragalactic Database (NED), which is operated by the Jet Propulsion Laboratory, California Institute of Technology, under contract with NASA. This research also made use of the Simbad and VizieR databases, operated at CDS, Strasbourg, France. We also acknowledge extensive use of the HPOL spectropolarimetric database,  ``http://www.sal.wisc.edu/HPOL/'' for instrument development and calibration. We further acknowledge the variable star observations from the AAVSO International Database contributed by observers worldwide and used in this research.

\appendix

\section{CPT-Even Q and U}
\label{sec:david}

We can calculate the Stokes Q and U parameters in the presence of CPT-even SME coefficients of the form $\kdEB$ as defined in Eq.~(\ref{eq:kdEB}). As was written in Eq.~(\ref{eq:PhiiEB}), the phase delay between the two normal modes is given by the equation
\begin{eqnarray}
\Phi(\omega) = 2 \omega^{d-3} L^{(d)} \kdEB\, ,
\end{eqnarray}
where we denote the energy as $\omega$ as opposed to $E$ in this case to distinguish it from the electric field.

The most conservative limits on SME coefficients are obtained when we assume a broadband source emitting a uniformly linearly polarized electric field along our line of sight $\hat z$ in the form
\begin{eqnarray}
\vec E(\omega, t) = Ae^{i\omega t} \hat n\, ,
\end{eqnarray}
where $\hat n$ makes an angle $\phi_0$. Then if the slow axis of this CPT-even Lorentz violation makes an angle $\phi_b$ so that we can define the quantity
\begin{eqnarray}
\Psi = \psi_0 - \psi_b\, ,
\end{eqnarray}
then the signal that reaches our detector along the slow and fast axes can be written
\begin{eqnarray}
E_{slow}(\omega, t) = Ae^{i\omega t - i\Phi/2} \cos\Psi\, , \\
E_{fast}(\omega, t) = Ae^{i\omega t + i\Phi/2} \sin\Psi\, ,
\end{eqnarray}
which, relative to our detector, is the electric field
\begin{eqnarray}
\vec E(\omega, t) & = & Ae^{i\omega t} \left[ e^{-i\Phi/2} \cos\Psi\cos\psi_b - e^{i\Phi/2} \sin\Psi\sin\phi_b \right] \hat x \nonumber \\
                  & + & Ae^{i\omega t} \left[ e^{-i\Phi/2} \cos\Psi\sin\psi_b + e^{i\Phi/2} \sin\Psi\cos\phi_b \right] \hat y\, . \nonumber \\
\end{eqnarray}

Next we can define averaging over the transmission band $T(\omega)$ as the operation
\begin{eqnarray}
\left<X\right>_\omega = \frac{\int d\omega T(\omega) X(\omega)}{\int d\omega T(\omega)}\, ,
\end{eqnarray}
so that the Stokes parameters in terms of the band averaged electric field $\vec E(t) = \left<\vec E(\omega,t )\right>_\omega$ incident on our detector are
\begin{eqnarray}
I & = & |E_x|^2 + |E_y|^2  = |A|^2\, , \\
Q & = & |E_x|^2 - |E_y|^2   \\ 
& = & I\cos2\Psi\cos2\psi_b-I|\left<\cos\Phi\right>|^2\sin2\Psi\sin\phi_b\, , \nonumber\\
U & = & 2\ \mathrm{Re} (E_x E_y^*)  \\
& = & I\cos2\Psi\sin2\psi_b+I|\left<\cos\Phi\right>|^2\sin2\Psi\cos\phi_b\, , \nonumber \\
V & = & 2\ \mathrm{Im} (E_x E_y^*)  = I|\left<\sin\Phi\right>|^2\sin2\Psi\, ,
\end{eqnarray}
therefore the normalized Stokes parameters $q=Q/I,u=U/I$ and total linear polarization fraction $p$ are
\begin{eqnarray}
q & = & \cos2\Psi\cos2\psi_b-|\left<\cos\Phi\right>|^2\sin2\Psi\sin\phi_b\, , \\
u & = & \cos2\Psi\sin2\psi_b+|\left<\cos\Phi\right>|^2\sin2\Psi\cos\phi_b\, , \\
p & = & \sqrt{\cos^2 2\Psi + \left<\cos\Phi\right>^2\sin^2 2\Psi}\, .
\label{eq:quCPTeven}
\end{eqnarray}
Due to the many unknowns in Eqs.~(\ref{eq:quCPTeven}), it is impractical to use time delays between each $q$ and $u$ time series, for example, to constrain SME vacuum birefringent parameters. Circular polarization measurements could potentially break certain degeneracies, but the maximum observed polarization approach will, in general, yield more sensitive SME constraints than any approaches using optical time delays.


\section{Data Plots and Tables}
\label{sec:data1}

\begin{figure*}
\centering
\begin{tabular}{@{}c@{}}

\includegraphics[width=7.2in]{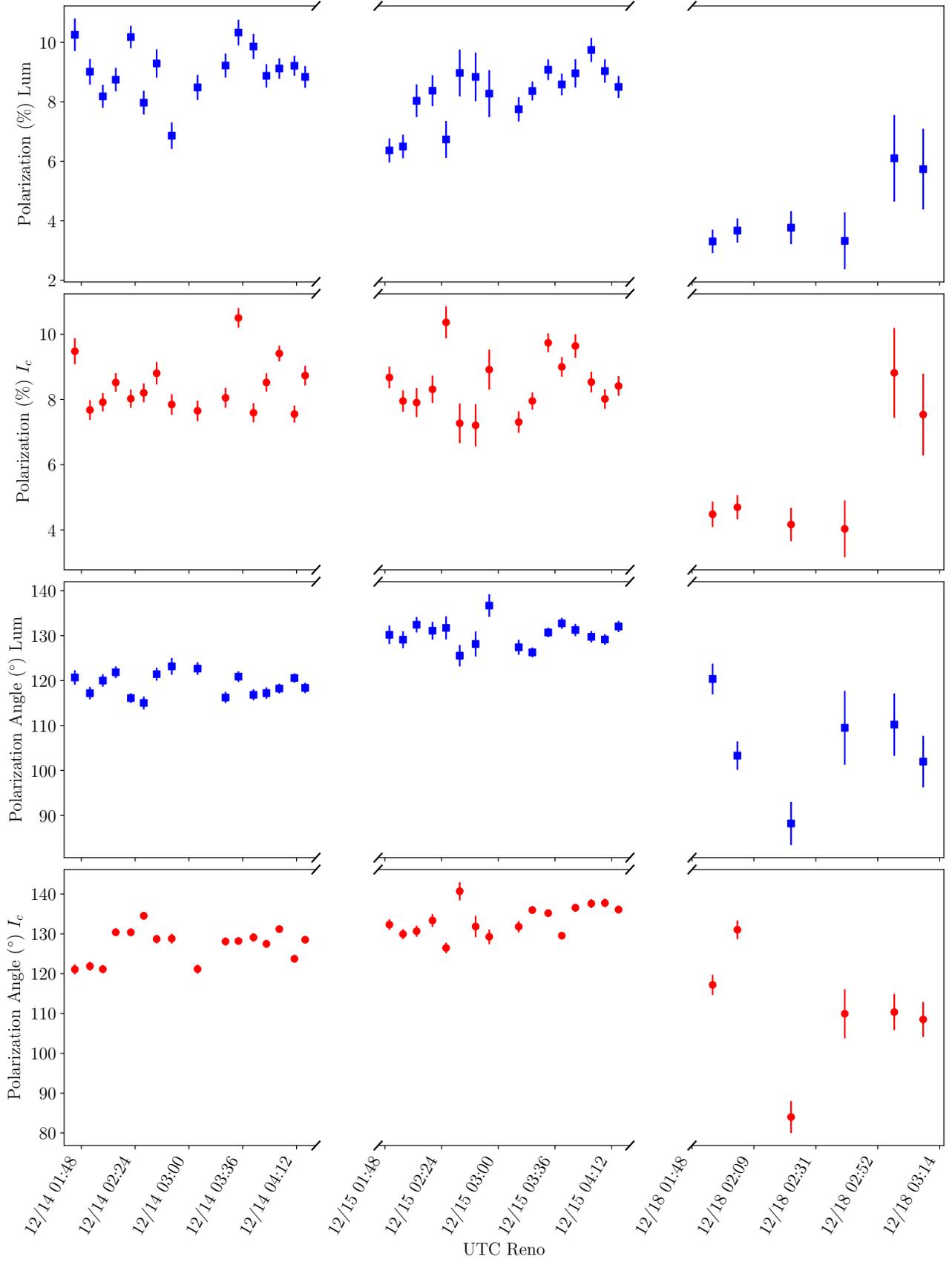} \\

\end{tabular}
\vspace{-0.6cm}
\caption{
\baselineskip 9pt 
Polarization $p$ (in \%) and polarization angle $\psi$ (in degrees) for BL Lacertae in the $Lum$ and $I_c$ bands.
}
\label{fig:bllac1a}
\end{figure*}

\begin{figure*}
\centering
\begin{tabular}{@{}c@{}}

\includegraphics[width=7.2in]{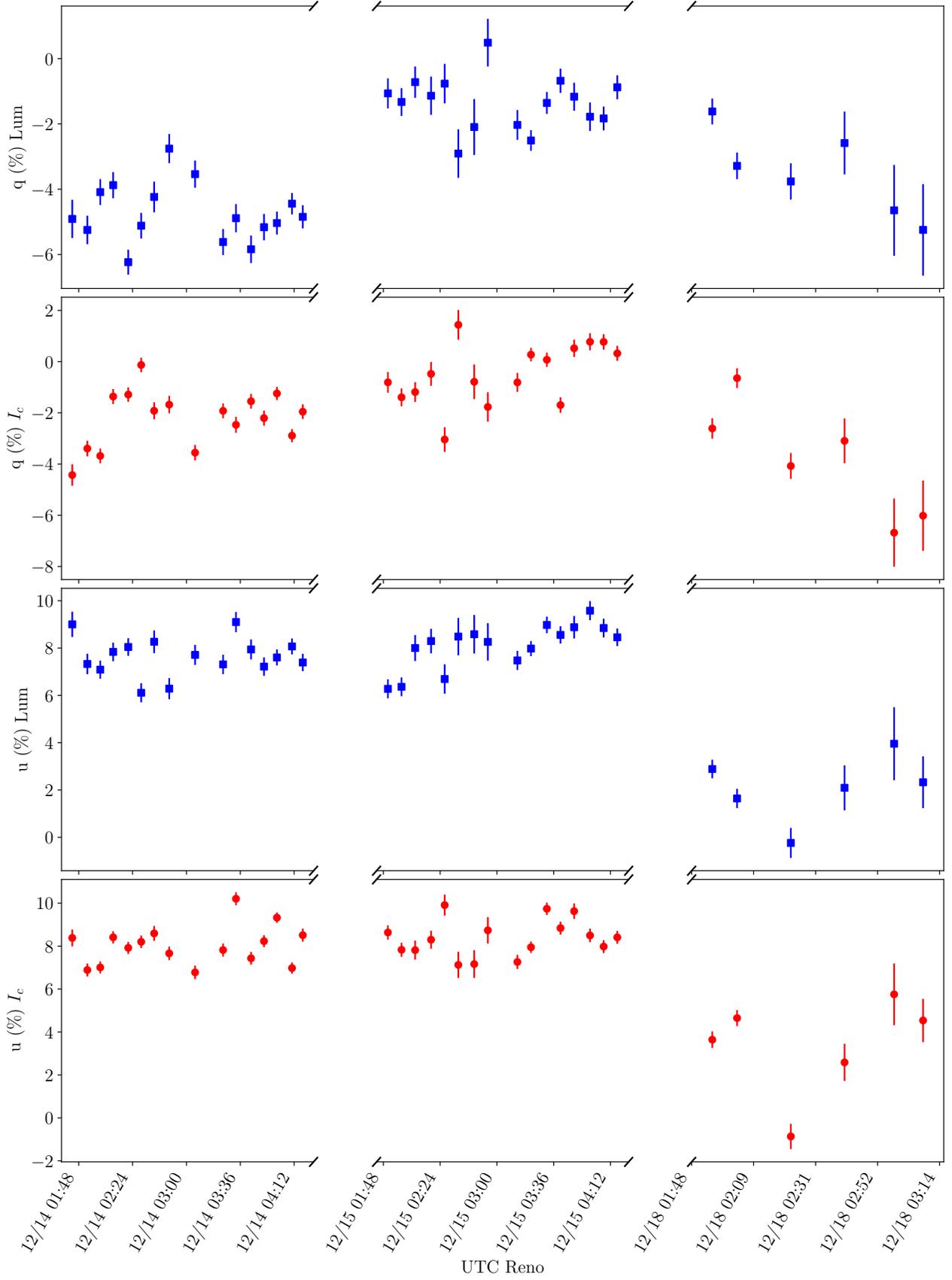} \\

\end{tabular}
\caption{
\baselineskip 9pt 
BL Lacertae polarimetric light curves for intensity normalized Stokes parameters $\Pq \equiv Q/I$ and $\Pu \equiv U/I$ in the $Lum$ and $I_c$ bands.
}
\label{fig:bllac1b}
\end{figure*}

\begin{figure*}
\centering
\begin{tabular}{@{}c@{}}

\includegraphics[width=7.2in]{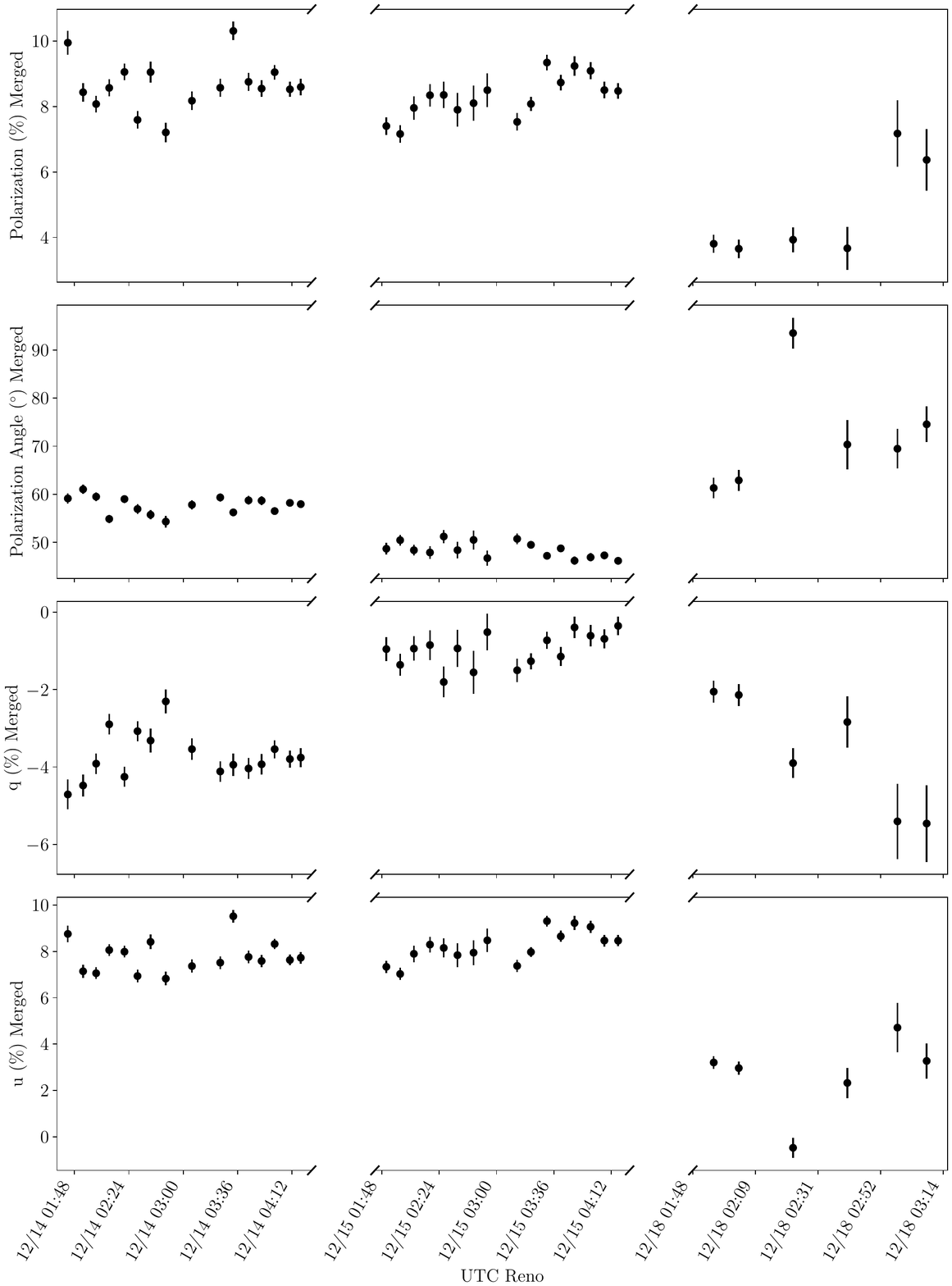} \\

\end{tabular}
\caption{
\baselineskip 9pt 
BL Lacertae light curves for polarization $p$ (in \%), polarization angle $\psi$ (in degrees), intensity normalized Stokes parameters $\Pq \equiv Q/I$, and $\Pu \equiv U/I$ in the {\it Merged} $Lum+I_c$ band.
}
\label{fig:bllac1c}
\end{figure*}

\begin{figure*}
\centering
\begin{tabular}{@{}c@{}}

\includegraphics[width=7.2in]{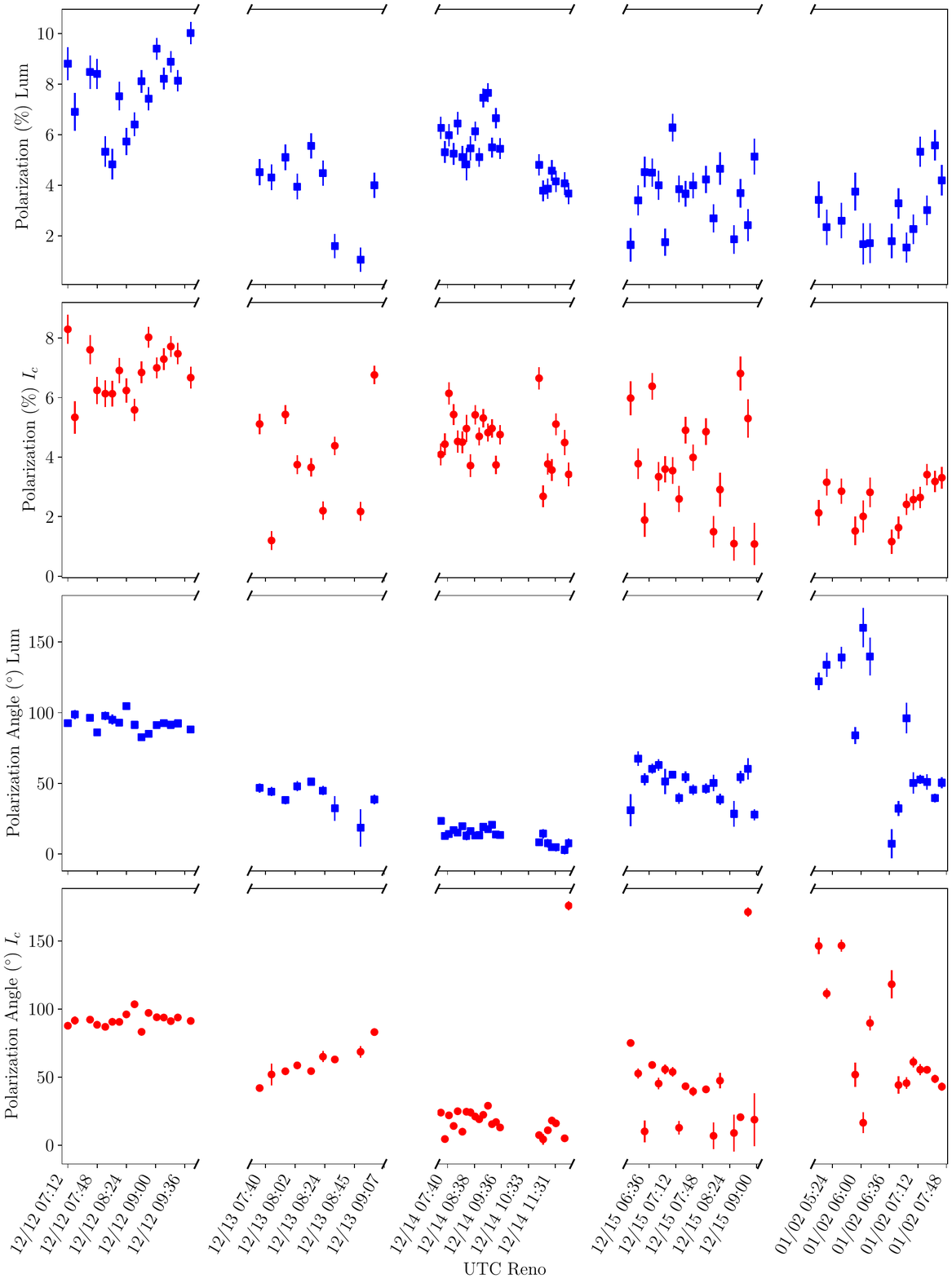} \\

\end{tabular}
\vspace{-0.6cm}
\caption{
\baselineskip 9pt 
Polarization $p$ (in \%) and polarization angle $\psi$ (in degrees) for \stwo{} in the $Lum$ and $I_c$ bands.
}
\label{fig:s5data1a}
\end{figure*}

\begin{figure*}
\centering
\begin{tabular}{@{}c@{}}

\includegraphics[width=7.2in]{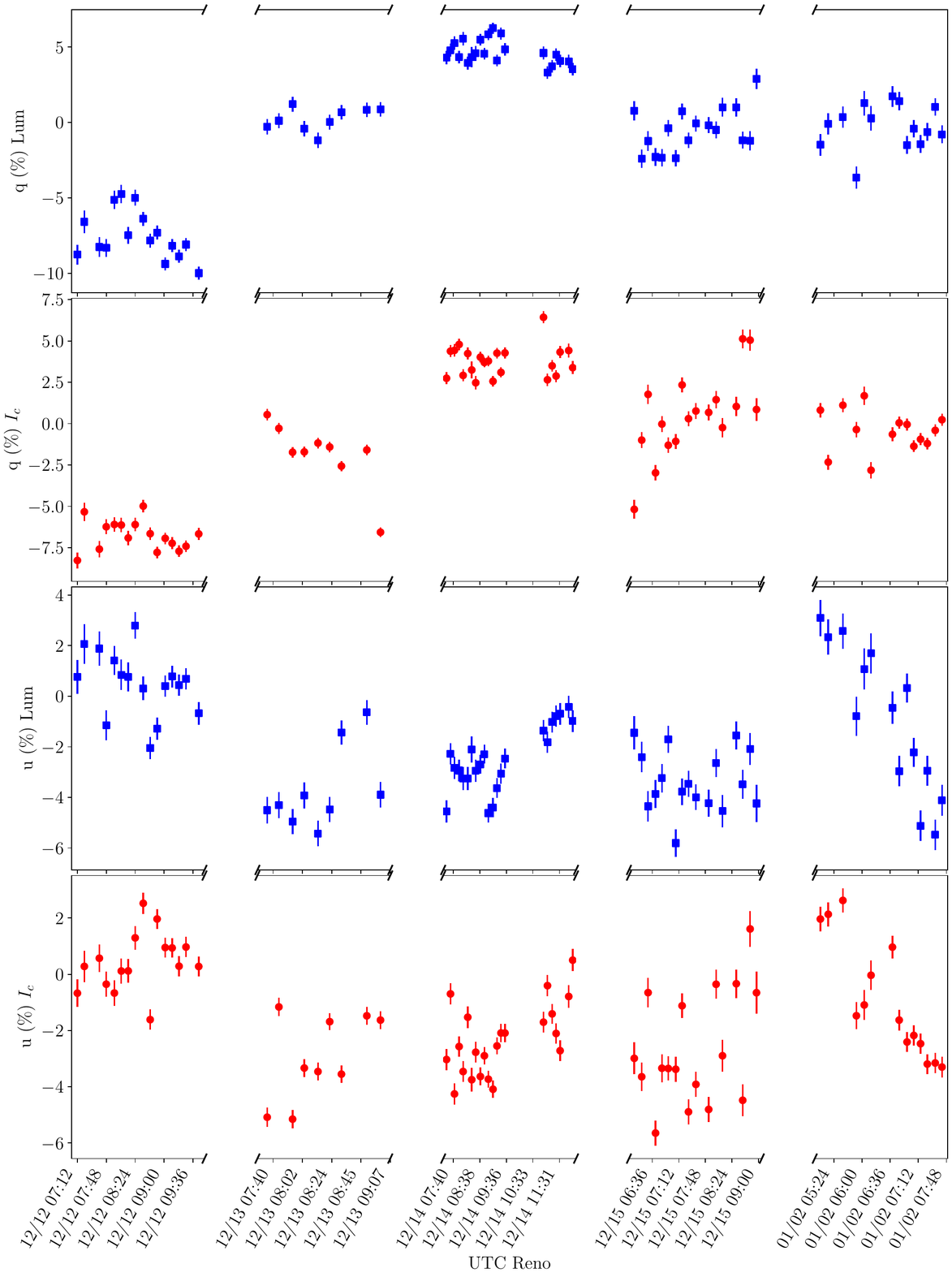} \\

\end{tabular}
\caption{
\baselineskip 9pt 
\stwo{} polarimetric light curves for intensity normalized Stokes parameters $\Pq \equiv Q/I$ and $\Pu \equiv U/I$ in the $Lum$ and $I_c$ bands.
}
\label{fig:s5data1b}
\end{figure*}

\begin{figure*}
\centering
\begin{tabular}{@{}c@{}}

\includegraphics[width=7.2in]{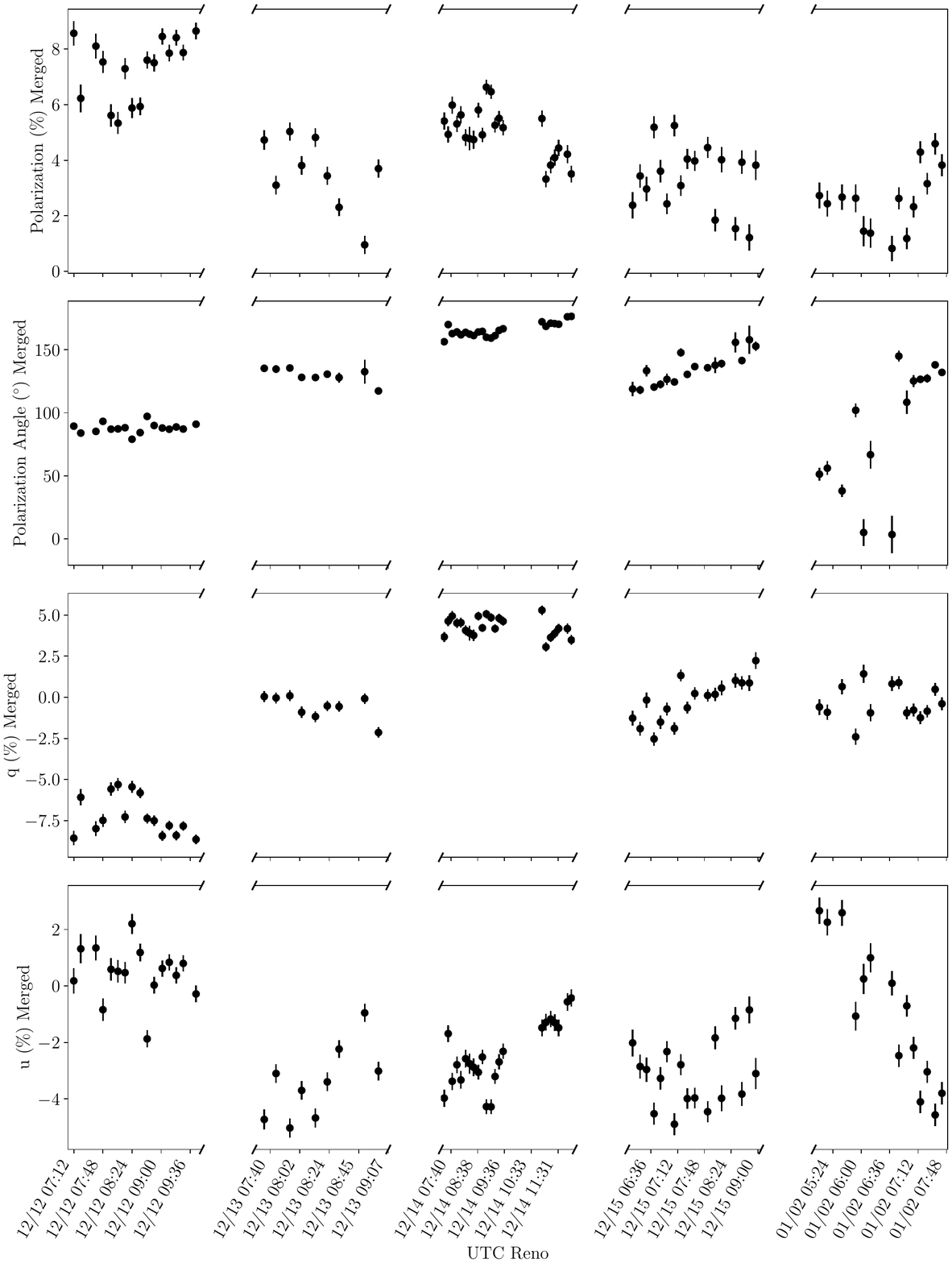} \\

\end{tabular}
\caption{
\baselineskip 9pt 
\stwo{} light curves for polarization $p$ (in \%), polarization angle $\psi$ (in degrees), intensity normalized Stokes parameters $\Pq \equiv Q/I$, and $\Pu \equiv U/I$ in the {\it Merged} $Lum+I_c$ band.
}
\label{fig:s5data1c}
\end{figure*}

\begin{figure*}
\centering
\begin{tabular}{@{}c@{}}

\includegraphics[width=7.2in]{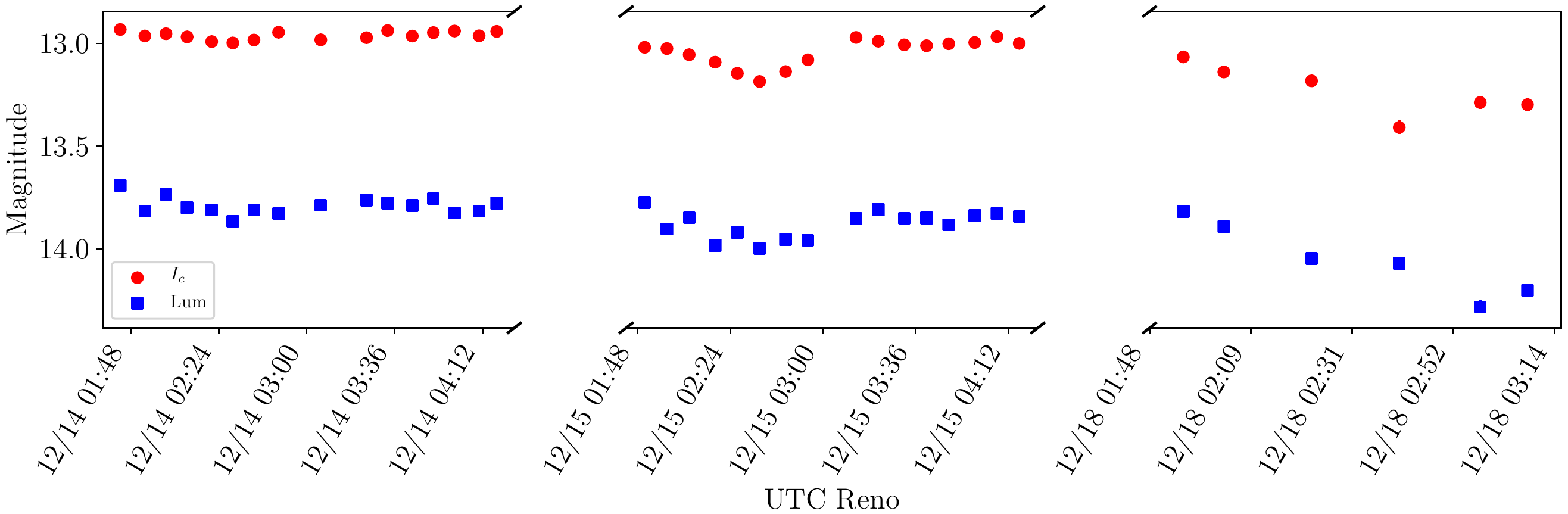} \\

\end{tabular}
\caption{
\baselineskip 9pt 
$I_c$ and $Lum$-band Photometry for BL Lacertae. 
}
\label{fig:bllac1p}
\end{figure*}

\begin{figure*}
\centering
\begin{tabular}{@{}c@{}}

\includegraphics[width=7.2in]{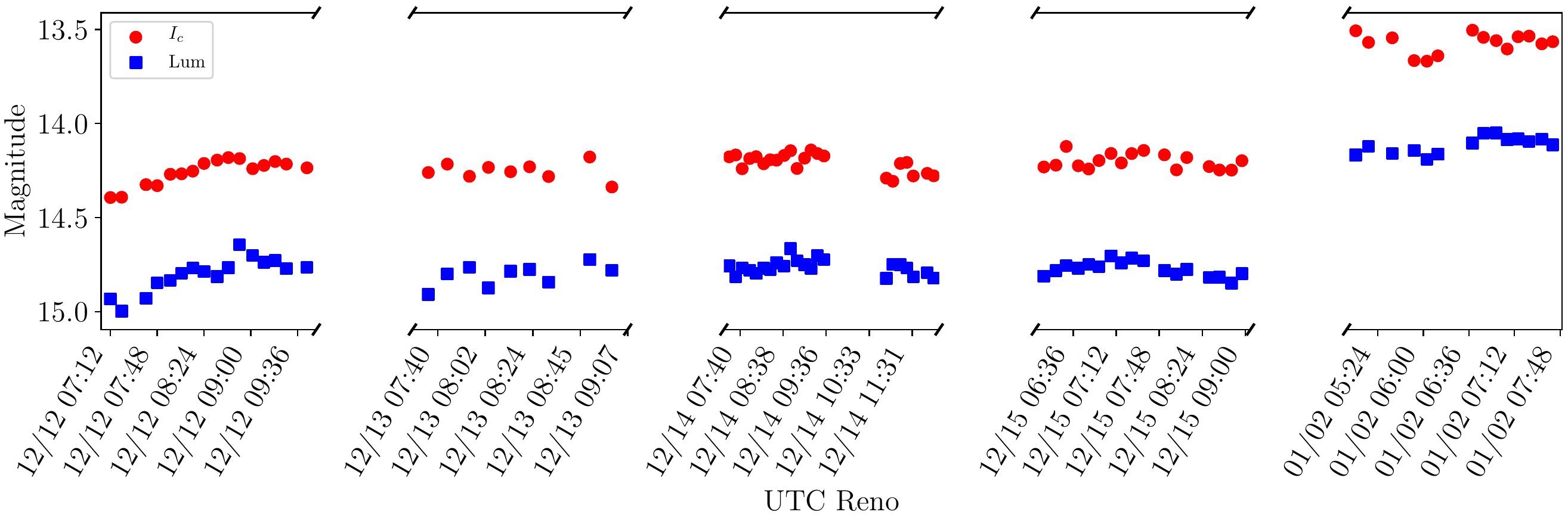} \\

\end{tabular}
\caption{
\baselineskip 9pt 
$I_c$ and $Lum$-band Photometry for \stwo{}. 
}
\label{fig:s5data1p}
\end{figure*}

All data in this paper were observed with \apol{} over December 2017 - January 2018. Data samples for BL Lacertae and \stwo{} are shown in Tables~\ref{tab:data1}-\ref{tab:data2a}, with the full machine-readable data to be made available online at \url{http://cosmology.ucsd.edu} and the journal website. Polarimetry is plotted in Figs.~\ref{fig:bllac1a}-\ref{fig:bllac1c} for BL Lacertae and in Figs.~\ref{fig:s5data1a}-\ref{fig:s5data1c} for \stwo{}, with photometry in Figs.~\ref{fig:bllac1p} and~\ref{fig:s5data1p}, respectively.


\begin{table*}[h]
\centering 
\footnotesize
\begin{tabular}{@{\cspace}l@{\cspace}r@{\cspace}r@{\cspace}r@{\cspace}r@{\cspace}r@{\cspace}r@{\cspace}r@{\cspace}r@{\cspace}r@{\cspace}r@{\cspace}} 
\hline 
  \multicolumn{1}{c}{MJD}   
& \multicolumn{1}{c}{$p_{L}$} 
& \multicolumn{1}{c}{$p_{I}$}    
& \multicolumn{1}{c}{$\psi_{L}$}  
& \multicolumn{1}{c}{$\psi_{I}$} 
& \multicolumn{1}{c}{$q_{L}$} 
& \multicolumn{1}{c}{$q_{I}$}    
& \multicolumn{1}{c}{$u_{L}$} 
& \multicolumn{1}{c}{$u_{I}$}
& \multicolumn{1}{c}{$m_{L}$}
& \multicolumn{1}{c}{$m_{I}$} \\ 
  \multicolumn{1}{c}{(days)}        
& \multicolumn{1}{c}{(\%)}   
& \multicolumn{1}{c}{(\%)}    
& \multicolumn{1}{c}{(deg)} 
& \multicolumn{1}{c}{(deg)} 
& \multicolumn{1}{c}{(\%)} 
& \multicolumn{1}{c}{(\%)} 
& \multicolumn{1}{c}{(\%)} 
& \multicolumn{1}{c}{(\%)} 
& \multicolumn{1}{c}{(mag)} 
& \multicolumn{1}{c}{(mag)} \\ 
\hline 
58101.072 & 10.3 $\pm$ 0.6 & 9.5 $\pm$ 0.4 & 59.0 $\pm$ 2.0 & 59.0 $\pm$ 1.0 & -4.9 $\pm$ 0.6 & -4.4 $\pm$ 0.4 & 9.0 $\pm$ 0.5 & 8.4 $\pm$ 0.4 & 13.69 $\pm$ 0.03 & 12.93 $\pm$ 0.04 \\ 
58101.079 & 9.0 $\pm$ 0.4 & 7.7 $\pm$ 0.3 & 63.0 $\pm$ 1.0 & 58.0 $\pm$ 1.0 & -5.2 $\pm$ 0.4 & -3.4 $\pm$ 0.3 & 7.3 $\pm$ 0.4 & 6.9 $\pm$ 0.3 & 13.82 $\pm$ 0.03 & 12.96 $\pm$ 0.04 \\ 
58101.085 & 8.2 $\pm$ 0.4 & 7.9 $\pm$ 0.3 & 60.0 $\pm$ 1.0 & 59.0 $\pm$ 1.0 & -4.1 $\pm$ 0.4 & -3.7 $\pm$ 0.3 & 7.1 $\pm$ 0.4 & 7.0 $\pm$ 0.3 & 13.74 $\pm$ 0.03 & 12.95 $\pm$ 0.04 \\ 
\hline 
\end{tabular} 
\caption{
Three nights of data for BL Lacertae were observed using \apol{} on December 13, 14, and 17, 2017. For each Modified Julian Date (MJD), we show our observed polarimetric and photometric data for both the $Lum$ and $I_c$-bands, denoted by $L$ and $I$ subscripts, respectively. Columns include the observed polarization $p$ (in \%), the polarization angle $\psi$ (in degrees), the intensity normalized Stokes parameters $q$ and $u$ (note that both can be negative, even when expressed as percentages), and the observed magnitude $m$. (A portion of this table is shown for guidance. This table is available in its entirety in machine-readable form.)
}
\label{tab:data1} 
\end{table*} 

\begin{table*}[h]
\centering 
\footnotesize
\begin{tabular}{@{\cspace}l@{\cspace}r@{\cspace}r@{\cspace}r@{\cspace}r@{\cspace}r@{\cspace}r@{\cspace}r@{\cspace}r@{\cspace}r@{\cspace}r@{\cspace}} 
\hline 
  \multicolumn{1}{c}{MJD}   
& \multicolumn{1}{c}{$p_{L}$} 
& \multicolumn{1}{c}{$p_{I}$}    
& \multicolumn{1}{c}{$\psi_{L}$}  
& \multicolumn{1}{c}{$\psi_{I}$} 
& \multicolumn{1}{c}{$q_{L}$} 
& \multicolumn{1}{c}{$q_{I}$}    
& \multicolumn{1}{c}{$u_{L}$} 
& \multicolumn{1}{c}{$u_{I}$}
& \multicolumn{1}{c}{$m_{L}$}
& \multicolumn{1}{c}{$m_{I}$} \\ 
  \multicolumn{1}{c}{(days)}        
& \multicolumn{1}{c}{(\%)}   
& \multicolumn{1}{c}{(\%)}    
& \multicolumn{1}{c}{(deg)} 
& \multicolumn{1}{c}{(deg)} 
& \multicolumn{1}{c}{(\%)} 
& \multicolumn{1}{c}{(\%)} 
& \multicolumn{1}{c}{(\%)} 
& \multicolumn{1}{c}{(\%)} 
& \multicolumn{1}{c}{(mag)} 
& \multicolumn{1}{c}{(mag)} \\ 
\hline 
58099.300 & 8.8 $\pm$ 0.7 & 8.3 $\pm$ 0.5 & 88.0 $\pm$ 2.0 & 92.0 $\pm$ 2.0 & -8.8 $\pm$ 0.7 & -8.3 $\pm$ 0.5 & 0.8 $\pm$ 0.7 & -0.7 $\pm$ 0.5 & 14.93 $\pm$ 0.04 & 14.39 $\pm$ 0.05 \\ 
58099.306 & 6.9 $\pm$ 0.7 & 5.3 $\pm$ 0.5 & 81.0 $\pm$ 3.0 & 88.0 $\pm$ 3.0 & -6.6 $\pm$ 0.7 & -5.3 $\pm$ 0.5 & 2.1 $\pm$ 0.8 & 0.3 $\pm$ 0.6 & 15.00 $\pm$ 0.04 & 14.39 $\pm$ 0.05 \\ 
58099.319 & 8.5 $\pm$ 0.7 & 7.6 $\pm$ 0.5 & 84.0 $\pm$ 2.0 & 88.0 $\pm$ 2.0 & -8.3 $\pm$ 0.7 & -7.6 $\pm$ 0.5 & 1.9 $\pm$ 0.7 & 0.6 $\pm$ 0.5 & 14.93 $\pm$ 0.04 & 14.32 $\pm$ 0.05 \\ 
\hline 
\end{tabular} 
\caption{
Same as Table~\ref{tab:data1} but all data for \stwo, for which we observed data using \apol{} over five nights on December 11-14 2017 and January 1, 2018. (A portion of this table is shown for guidance. This table is available in its entirety in machine-readable form.)
}
\label{tab:data2} 
\end{table*} 

\begin{table*}[h]
\centering 
\footnotesize
\begin{tabular}{@{\cspace}l@{\cspace}r@{\cspace}r@{\cspace}r@{\cspace}r@{\cspace}} 
\hline 
  \multicolumn{1}{c}{MJD}               
& \multicolumn{1}{c}{$p_{L+I}$}    
& \multicolumn{1}{c}{$\psi_{L+I}$}  
& \multicolumn{1}{c}{$q_{L+I}$} 
& \multicolumn{1}{c}{$u_{L+I}$} \\ 
  \multicolumn{1}{c}{(days)}        
& \multicolumn{1}{c}{(\%)}   
& \multicolumn{1}{c}{(deg)} 
& \multicolumn{1}{c}{(\%)} 
& \multicolumn{1}{c}{(\%)} \\ 
\hline 
58101.072 & 9.9 $\pm$ 0.4 & 59.0 $\pm$ 1.0 & -4.7 $\pm$ 0.4 & 8.8 $\pm$ 0.4 \\ 
58101.079 & 8.4 $\pm$ 0.3 & 61.0 $\pm$ 1.0 & -4.5 $\pm$ 0.3 & 7.1 $\pm$ 0.3 \\ 
58101.085 & 8.1 $\pm$ 0.3 & 59.5 $\pm$ 0.9 & -3.9 $\pm$ 0.3 & 7.1 $\pm$ 0.3 \\ 
\hline 
\end{tabular} 
\caption{
Same as Table~\ref{tab:data1} for BL Lacertae, but for the combined $Lum+I_c$-band. (A portion of this table is shown for guidance. This table is available in its entirety in machine-readable form.)
}
\label{tab:data1a} 
\end{table*} 

\begin{table*}[h]
\centering 
\footnotesize
\begin{tabular}{@{\cspace}l@{\cspace}r@{\cspace}r@{\cspace}r@{\cspace}r@{\cspace}} 
\hline 
  \multicolumn{1}{c}{MJD}               
& \multicolumn{1}{c}{$p_{L+I}$}    
& \multicolumn{1}{c}{$\psi_{L+I}$}  
& \multicolumn{1}{c}{$q_{L+I}$} 
& \multicolumn{1}{c}{$u_{L+I}$} \\ 
  \multicolumn{1}{c}{(days)}        
& \multicolumn{1}{c}{(\%)}   
& \multicolumn{1}{c}{(deg)} 
& \multicolumn{1}{c}{(\%)} 
& \multicolumn{1}{c}{(\%)} \\ 
\hline 
58099.300 & 8.6 $\pm$ 0.4 & 89.0 $\pm$ 1.0 & -8.6 $\pm$ 0.4 & 0.2 $\pm$ 0.4 \\ 
58099.306 & 6.2 $\pm$ 0.5 & 84.0 $\pm$ 2.0 & -6.1 $\pm$ 0.5 & 1.3 $\pm$ 0.5 \\ 
58099.319 & 8.1 $\pm$ 0.4 & 85.0 $\pm$ 2.0 & -8.0 $\pm$ 0.4 & 1.3 $\pm$ 0.4 \\ 
\hline 
\end{tabular} 
\caption{
Same as Table~\ref{tab:data2} for \stwo, but for the combined $Lum+I_c$-band. 
(A portion of this table is shown for guidance. This table is available in its entirety in machine-readable form.)
}
\label{tab:data2a} 
\end{table*} 

\clearpage


%


\end{document}